\documentclass[pra,aps,twocolumn,longbibliography]{revtex4-1}
\usepackage{graphicx,amsmath,bm}
\usepackage{txfonts}
\usepackage{color}
\usepackage{hyperref}
\newcommand{\rp}[1]{(\ref{#1})}

\newcommand{\abs}[1]{\left|{#1}\right|}

\newcommand{\av}[1]{\left\langle #1 \right\rangle}

\newcommand{\al}[1]{^{(#1)}}
\newcommand{\da}{^\dagger}

\newcommand{\pt}[1]{\left( #1 \right)}
\newcommand{\pq}[1]{\left[ #1 \right]}
\newcommand{\pg}[1]{\left\{ #1 \right\}}

\newcommand{\lpg}[1]{\left\{ #1 \right.}

\newcommand{\rpg}[1]{\left. #1 \right\}}
\newcommand{\ee}{{\rm e}}
\newcommand{\ii}{{\rm i}}
\newcommand{\dd}{{\rm d}}

\newcommand{\pd}[1]{\frac{\partial}{\partial #1}}

\newcommand{\nn}{{\nonumber}}

\newcommand{\LL}{{\cal L}}

\newcommand{\mm}{{\rm m}}
\newcommand{\fb}{{\rm fb}}

\newcommand{\eff}{{\rm eff}}
\newcommand{\out}{{\rm out}}
\newcommand{\inn}{{\rm in}}

\newcommand{\T}{{\rm T}}

\usepackage[normalem]{ulem}  
\begin{document}

\title{
Optomechanical Stirling heat engine driven by feedback-controlled light
}

\author{Giacomo Serafini}
\author{Stefano Zippilli}
\author{Irene Marzoli}
\affiliation{School of Science and Technology, Physics Division, University of Camerino, I-62032 Camerino (MC), Italy}

\date{\today}

\begin{abstract}
We propose and analyze a microscopic Stirling heat engine based on an optomechanical system.
The working fluid is a single vibrational mode of a mechanical resonator, which interacts by radiation pressure with a feedback-controlled optical cavity.
The cavity light is used to engineer the thermal reservoirs and to steer the resonator through a thermodynamic cycle.
In particular, the feedback is used to properly modulate the light fluctuations inside the cavity and hence to realize efficient thermodynamic transformations with realistic optomechanical devices. 
\end{abstract}
\maketitle

% \tableofcontents
% \makeatletter
% \let\toc@pre\relax
% \let\toc@post\relax
% \makeatother

\section{Introduction}

Microscopic thermal machines are intensively employed in the investigation of the thermodynamics of the quantum world and of non-equilibrium systems~\cite{benenti2017,binder2018}.
Various proposals exploit optomechanical systems, which are very versatile devices that can be controlled also at the quantum level~\cite{aspelmeyer2014,bowen2015}, to design microscopic heat engines~\cite{zhang2014a,zhang2014,dong2015,dong2015a,dechant2015a,mari2015,gelbwaser-klimovsky2015a,Bathaee2016,zhang2017,bennett2020a,naseem2019,abari2019}. However no experimental optomechanical heat engine has been demonstrated so far. Here we propose and analyze a feasible heat engine based on a feedback-controlled optomechanical system~\cite{zippilli2018}.

Specifically, in this work we study an optomechanical system within a feedback loop, where the amplitude of the laser field, which drives the optical cavity, is modulated by a signal, 
proportional to the photocurrent resulting from the homodyne detection of a field quadrature at the cavity output (see Fig.~\ref{Fig1}). 
This scheme is similar to the one investigated in Refs.~\cite{rossi2017,Kralj2017,rossi2018,zippilli2018,abari2019} and the present research is a further demonstration of the versatility of feedback-controlled 
light, as an efficient tool to steer the dynamics of optomechanical systems.
Here we show, analogously to Ref.~\cite{abari2019}, that by controlling the light fluctuations with a feedback system it is possible to engineer an efficient optomechanical-based heat engine. 
However, the dynamics that we describe in this work is fundamentally different from that reported in Ref.~\cite{abari2019}.
Here the engine employs the phononic excitations of the mechanical resonator as the working fluid. 
Laser cooling is used to engineer the thermal baths at different temperatures, with which the resonator comes into contact during the thermodynamic cycle.
Moreover, the optical spring effect is exploited to modulate the mechanical frequency, which mimics the variation of the volume in a standard thermodynamic heat engine. 
Differently from typical optomechanical systems, where the optical spring is relatively small to achieve a sizable efficiency in an engine of this kind, 
here we employ the feedback system to increase the optical spring and attain a significant efficiency. 
The resulting dynamics is somehow similar to that investigated in Ref.~\cite{dechant2015a}, which describes how to realize an optomechanical Stirling heat engine using a trapped nanoparticle. 
In this latter case, a sufficiently large variation of the mechanical frequency, i.e., the trapped particle oscillation frequency, is realized by  controlling the intensity of the additional trapping
potential.
We highlight that our approach makes use of a simpler optical set-up, which can be applied to any optomechanical device -- not only trapped particles -- both in the optical and microwave domain.

Moreover, it is important to emphasize that this setup can be potentially exploited to investigate the role of correlations in the energy exchanges of microscopic and quantum heat engines~\cite{gelbwaser-klimovsky2015b,uzdin2016,wiedmann2020a,newman2020,abah2014a,niedenzu2016,campisi2017,potts2018}. In fact the feedback scheme that we consider can be exploited also to engineer and control both the correlations in the effective mechanical reservoir and between the mechanical resonator and the reservoir~\cite{zippilli2018}. 

The article is organized as follows. In Sec.~\ref{Model} we introduce the model and derive the effective reduced equations for the mechanical resonator, after the adiabatic elimination of the cavity mode. 
Then, in Sec.~\ref{Engine}, we identify a suitable Stirling cycle and show how to steer the system dynamics, by adjusting the feedback parameters. 
In Sec.~\ref{Efficiency} we study the engine performances under different conditions. Finally Sec.~\ref{Conclusions} is devoted to the conclusions. In the appendices we report the details of the adiabatic elimination of the cavity field (App.~\ref{App-ad-el}), as well as additional results, evaluated for systems in the resolved sideband regime (App.~\ref{App-rsl}).
 
\section{The Model}\label{Model}

\begin{figure}[t!]
\centering
\includegraphics[width=\columnwidth]{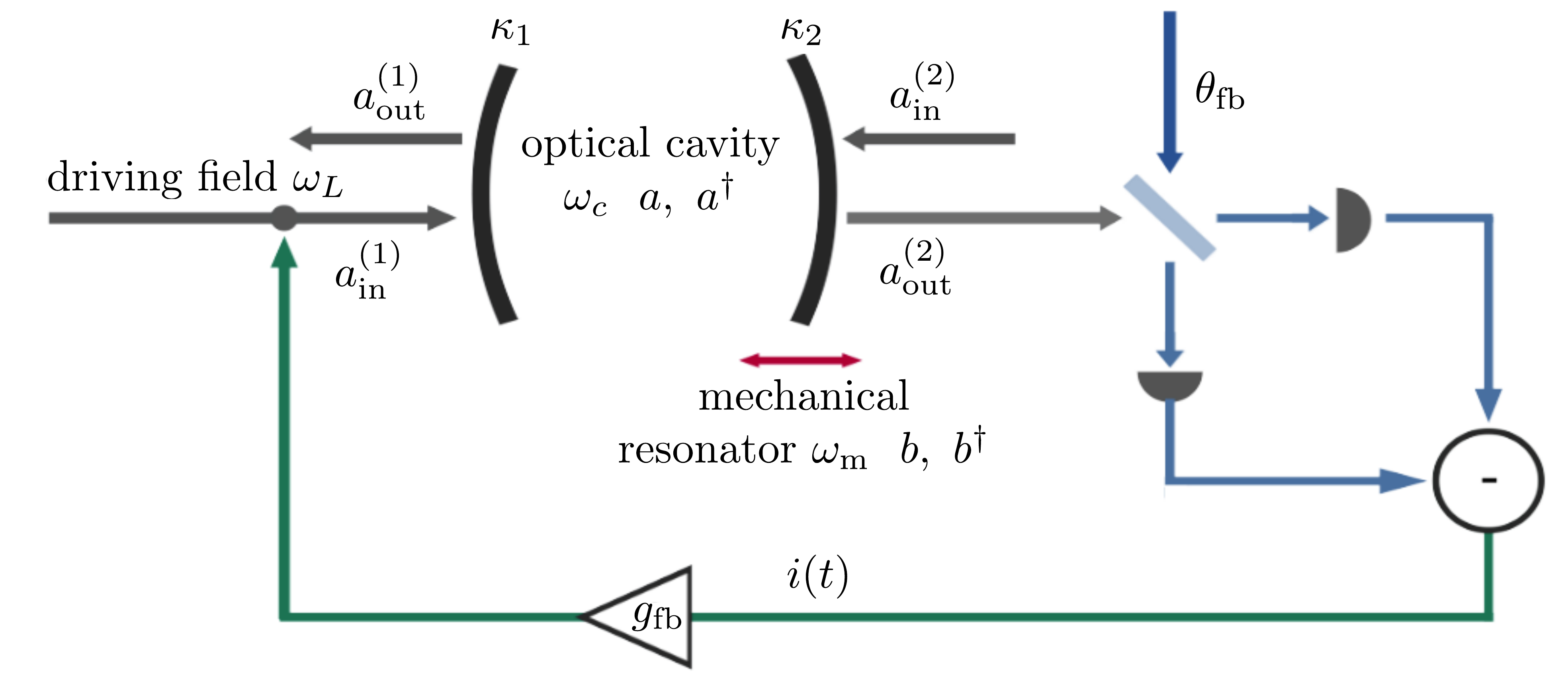}
\caption{Schematic drawing of the optomechanical system made of a Fabry-P\'erot optical cavity with a moving end mirror. The light transmitted through the optical cavity is measured by balanced homodyne detection, with $\theta_\fb$
being the phase difference between the signal and the local oscillator.
The corresponding photocurrent $i(t)$ is used to modulate the input amplitude via the feedback loop.}
\label{Fig1}
\end{figure}
We consider a mode of a Fabry-P\'erot optical cavity at the frequency $\omega_c$, driven by a laser field at the frequency $\omega_L$, and with total decay rate 
$\kappa \equiv \kappa_1+\kappa_2$, where $\kappa_1$  and $\kappa_2$ are the decay rates due to the mirror losses.
The applied driving field is detuned by $\Delta \equiv \omega_c-\omega_L$ from the cavity resonance. 
The laser amplitude is modulated by a signal proportional to the homodyne photocurrent of the field quadrature detected at the cavity output (see Fig.~\ref{Fig1}). 
The cavity mode is coupled by radiation pressure to a vibrational mode of a mechanical resonator with frequency $\omega_\mm$ and dissipation rate $\gamma$.
We employ the standard linearized description of the optomechanical dynamics~\cite{bowen2015}, which  -- under the assumption of a sufficiently large driving power --
focuses on the fluctuations of the optical, $a(t)$ and $a\da(t)$, and mechanical, $b(t)$ and $b\da(t)$, field operators around the corresponding average values. 
Here, as in Refs.~\cite{rossi2017,Kralj2017,rossi2018,zippilli2018,abari2019}, we assume a feedback transfer function ($g_\fb$ in Fig.~\ref{Fig1}), which realizes a high pass filter, such that the feedback does not act at low frequencies. This way the average optomechanical variables remain unaffected by the feedback.
The fluctuations, instead, fulfill the quantum Langevin equations
\begin{eqnarray}\label{dota}
\dot a(t)&=&-\pt{\kappa+\ii\,\Delta}a(t)-\ii\,G\pq{b(t)+b\da(t)}+\sqrt{2\,\kappa}\,a_\inn(t) \,,
\\
\dot b(t)&=&-\pt{\frac{\gamma}{2}+\ii\,\omega_\mm}b(t)-\ii\,G\pq{a(t)+a\da(t)}+\sqrt{\gamma}\,b_\inn(t) \,, 
\label{dotb}
\end{eqnarray}
where $G$ is the linearized coupling strength, proportional to the intensity of the cavity field.
The cavity detuning takes into account also the shift due to the optomechanical interaction. 
Moreover $a_\inn(t)$ and $b_\inn(t)$ are the input noise operators. 
The cavity input $a_\inn(t)$ includes the noise entering from the two mirrors 
\begin{eqnarray}\label{ain}
a_\inn(t) \equiv \frac{\sqrt{\kappa_1}\,a_\inn\al{1}(t)+\sqrt{\kappa_2}\,a_\inn\al{2}(t)}{\sqrt{\kappa}} \,.
\end{eqnarray}
The input field through the first mirror is modified -- assuming a broadband feedback response function~\cite{zippilli2018} -- according to the relation
\begin{eqnarray}\label{ain1}
a_\inn\al{1}(t)=a_{\inn,0}\al{1}(t)+g_\fb\ i(t-\tau_\fb)\ ,
\end{eqnarray}
where $g_\fb$ is the feedback gain and $i(t)$ is the homodyne photocurrent [defined below in Eq.~(\ref{photocurrent})], which is delayed by the feedback delay time $\tau_\fb$.
Note that the amplitude modulation described by Eq.~\rp{ain1} can be realized, as in Refs.~\cite{rossi2017,Kralj2017,rossi2018}, using an acousto-optic modulator driven by the detected photocurrent.
Furthermore $a_{\inn,0}\al{1}(t)$ is the input field without feedback and is characterized, as well as the input on the second mirror, by the vacuum noise fluctuations 
such that $\av{a_{\inn,0}\al{1}(t)\ a_{\inn,0}\al{1}{}\da(t')}=\av{a_{\inn}\al{2}(t)\ a_{\inn}\al{2}{}\da(t')}=\delta(t-t')$. 
The mechanical input noise, instead, describes thermal noise with $n_\T$ excitations such that $\av{b_{\inn}(t)\ b_{\inn}\da(t')}=\pt{n_\T+1}\delta(t-t')$.
Here, the feedback acts by measuring the optical field at the output of the second mirror.
After introducing the corresponding output operator 
\begin{eqnarray}
a_\out\al{2}(t)=\sqrt{2\,\kappa_2}\,a(t)-a_\inn\al{2}(t)\ ,
\end{eqnarray}
the homodyne photocurrent, with homodyne phase $\theta_\fb$
(that is the phase difference between the signal and the local oscillator)
and detection efficiency $\eta_d$, is
\begin{equation} \label{photocurrent}
i(t) = \sqrt{\eta_d}\pq{\ee^{-\ii\,\theta_\fb}\ a_\out\al{2}(t)+\ee^{\ii\,\theta_\fb}\ a_\out\al{2}{}\da(t) }+\sqrt{1-\eta_d}\ X_v(t) \,,
\end{equation}
where $X_v$ describes additional white noise, with correlation $\av{ X_v(t)\ X_v(t') }=\delta(t-t') $, due to the imperfect detection. 
Whenever the feedback delay time is much smaller than both the characteristic interaction time $1/G$ and the cavity decay time $1/\kappa$, 
its effect on the field operator amounts to an additional phase factor such that $a_\out\al{2}(t-\tau_\fb)\simeq a_\out\al{2}(t)\ \ee^{\ii\,\Delta\,\tau_\fb}$~\cite{abari2019}. Thereby we find 
\begin{eqnarray}\label{i}
i(t-\tau_\fb)\simeq\sqrt{\eta_d}\pq{\ee^{-\ii\,\phi}\ a_\out\al{2}(t)+\ee^{\ii\,\phi}\ a_\out\al{2}{}\da(t) }+\sqrt{1-\eta_d}\ X_v(t) \,,
\nn\\
\end{eqnarray}
with the total feedback phase denoted as 
\begin{equation}  \label{fb_phase}
       \phi \equiv \theta_\fb-\Delta\,\tau_\fb .
\end{equation}
By inserting Eqs.~\rp{ain}-\rp{i} into Eq.~(\ref{dota}), one finds that the Langevin equation for the cavity field takes the form
\begin{eqnarray}\label{dotaeff}
\dot a(t)&=&-\pt{\kappa_\eff+\ii\,\Delta_\eff}a(t)+
\mu^*\,a\da(t)
\nn\\&&
-\ii\,G\pq{b(t)+b\da(t)}+\sqrt{2\,\kappa_\eff}\ a_{\inn,\eff}(t) \,,
\end{eqnarray}
where we have introduced the feedback-modified parameters
\begin{eqnarray}\label{kappaeff-Deltaeff}
\kappa_\eff& \equiv &\kappa-{\rm Re}\pq{\mu} \,,
%2\sqrt{\eta_d\ \kappa_1\ \kappa_2}\ g_\fb\, \cos(\phi)
\nn\\
\Delta_\eff& \equiv &\Delta-{\rm Im}\pq{\mu} \,,
%+2\sqrt{\eta_d\ \kappa_1\ \kappa_2}\ g_\fb\, \sin(\phi)
\end{eqnarray}
with
\begin{eqnarray}\label{mu}
\mu& \equiv &2\sqrt{\eta_d\,\kappa_1\,\kappa_2}\,g_\fb\,\ee^{-\ii\phi}=\kappa-\kappa_\eff+\ii(\Delta-\Delta_\eff)\ .
\end{eqnarray}
Equations~\rp{kappaeff-Deltaeff} and \rp{mu} show that by controlling the feedback gain $g_\fb$  and phase $\phi$, it is possible to tune the effective response of the optical cavity~\cite{rossi2017,Kralj2017,rossi2018,zippilli2018}. For example, for positive feedback, i.e., ${\rm Re}\pq{\mu}>0$, the cavity linewidth can be effectively reduced. This effect can be exploited, as demonstrated in Ref.~\cite{rossi2017}, to achieve enhanced cooling of the mechanical resonator.

The total noise operator, which also accounts for the noise due to the feedback, is defined as 
\begin{eqnarray}
a_{\inn,\eff}(t)& \equiv &\frac{1}{\sqrt{\kappa_\eff}}\lpg{
\sqrt{\kappa_1}\ a_{\inn,0}\al{1}(t)+\sqrt{\kappa_2}\ a_\inn\al{2}(t)
}\nn\\&&
-\sqrt{\eta_d\ \kappa_1}\ g_\fb\pq{\ee^{-\ii\,\phi}\ a_\inn\al{2}(t)+\ee^{\ii\,\phi}\ a_\inn\al{2}{}\da(t) }
\nn\\&&\rpg{
+\sqrt{\pt{1-\eta_d}\kappa_1}\ g_\fb\ X_v(t)
}\ ,
\end{eqnarray}
with correlations
\begin{eqnarray}
\av{a_{\inn,\eff}(t)\ a_{\inn,\eff}\da(t')}&=&\pt{1+n_\eff}\delta(t-t') \,,
\\
\av{a_{\inn,\eff}(t)\ a_{\inn,\eff}(t')}&=&m_\eff\,\delta(t-t')\ ,
\label{aineff_corr}
\end{eqnarray}
where 
\begin{eqnarray}
n_\eff& \equiv &\frac{\kappa_1\ g_\fb^2}{\kappa_\eff}=\frac{\abs{\mu}^2}{4\,\eta_d\,\kappa_\eff\,\kappa_2}=\frac{\pt{\kappa-\kappa_\eff}^2+\pt{\Delta-\Delta_\eff}^2}{4\,\eta_d\,\kappa_\eff\,\kappa_2}\,,
\nn\\
m_\eff& \equiv &n_\eff\pt{1-\frac{\sqrt{\eta_d\,\kappa_2}\ \ee^{\ii\,\phi}}{\sqrt{\kappa_1}\ g_\fb}}=n_\eff\pt{1-\frac{2\,\eta_d\,\kappa_2}{\mu}}\ .
\end{eqnarray}
The feedback modifies the correlation properties of the input field. Now the input light field is no longer characterised by vacuum noise fluctuations, but exhibits a mean number of excitations ($n_\eff$) as well as self-correlations ($m_\eff$).

%%%%%%%%%%%%%%%%%%%%%%%%%%%%%%%%%%%%%%%

We are interested in the weak coupling regime $G\lesssim\kappa_\eff$, whereby the effect of the cavity light on the mechanical resonator can be taken into account by means of effective parameters, determined by adiabatically eliminating the cavity field from the Langevin equations of the mechanical resonator. 
Specifically we focus on the slowly varying mechanical operator $\bar b(t)$ defined by the transformation $b(t) \equiv \ee^{-\ii\,\omega_\mm t}\ \bar b(t)$, 
in comparison to which the cavity field evolves on a much shorter time scale. 
The resulting reduced equation for the mechanical degrees of freedom is (for details see App.~\ref{App-ad-el})
\begin{eqnarray}\label{dotb_adel}
\dot{\bar b}(t)& \simeq &-\pt{\frac{\gamma+\Gamma_\mm}{2}+\ii\Delta_\mm}\,\bar b(t)+\sqrt{\gamma+\Gamma_\mm}\,B_\inn(t) \,,
\end{eqnarray}
where we have introduced the light-induced mechanical dissipation rate $\Gamma_\mm$ 
and 
the optical spring $\Delta_\mm$,
which can be expressed in terms of the feedback-modified cavity response function
\begin{eqnarray}\label{Lambda}
\Lambda(\omega) \equiv \frac{\tilde\chi(\omega)\pq{1+\mu\,\tilde\chi(-\omega)^*}}{1-\abs{\mu}^2\,\tilde\chi(\omega)\,\tilde\chi(-\omega)^*}\ ,
\end{eqnarray}
where
\begin{eqnarray}\label{chi}
\tilde\chi(\omega) \equiv \frac{1}{\kappa_\eff+\ii\pt{\Delta_\eff-\omega}}\ ,
\end{eqnarray}
as
\begin{eqnarray}\label{Gammam}
\Gamma_\mm& \equiv &2\,G^2{\rm Re}\pq{\Lambda(\omega_\mm) - \Lambda(-\omega_\mm)^*}\,,
\\
\Delta_\mm& \equiv &G^2{\rm Im}\pq{\Lambda(\omega_\mm) - \Lambda(-\omega_\mm)^*}\ .
\label{Deltam}
\end{eqnarray}
Moreover, $B_\inn(t)$ is
the modified noise operator, whose correlation functions can be expressed in terms of the power spectrum of the cavity field amplitude (see App.~\ref{power-spectrum})
\begin{eqnarray}\label{SX0}
S_{X_0}(\omega)&=&2\,\kappa_\eff\lpg{ 
(n_\eff+1)\  \abs{\Lambda(\omega)}^2+  n_\eff \abs{\Lambda(-\omega)}^2 
}\nn\\&&\rpg{
+2{\rm Re}\pq{m_\eff\ \Lambda(\omega)^{\phantom{|}} \Lambda(-\omega)}
}\ ,
\end{eqnarray}
as
\begin{eqnarray}\label{BBcorrelations}
\av{B_\inn\da(t)\ B_\inn(t')}&\simeq&\frac{G^2\ S_{X_0}(-\omega_\mm)+\gamma\,n_\T}{\gamma+\Gamma_\mm}\ \delta(t-t') \,,
\nn\\
\av{B_\inn(t)\ B_\inn\da(t')}&\simeq&\frac{G^2\ S_{X_0}(\omega_\mm)+\gamma\,\pt{n_\T+1}}{\gamma+\Gamma_\mm}\ \delta(t-t') ,
\end{eqnarray}
and $\av{B_\inn(t)\ B_\inn(t')}=\av{B_\inn\da(t)\ B_\inn\da(t')}\simeq0$.
Hence, the corresponding steady state number of mechanical excitations, which describes the final stage of the laser cooling with feedback-controlled light~\cite{rossi2017,zippilli2018}, is
\begin{eqnarray}\label{nm}
n_\mm=\frac{G^2\ S_{X_0}(-\omega_\mm)+\gamma\,n_\T}{\gamma+\Gamma_\mm}\ .
\end{eqnarray}
Finally, we can introduce the corresponding equation for the average number of mechanical excitations, $n_b(t) \equiv \av{\bar b\da(t)\ \bar b(t)}$,
\begin{eqnarray} \label{ndot}
\dot n_b(t)=-\pt{\gamma+\Gamma_\mm}\pq{n_b(t)-n_\mm}\ ,
\end{eqnarray} 
which we are going to use in the evaluation of the engine performance. 

\begin{figure}[t!]
%\centering
\includegraphics[width=\columnwidth]{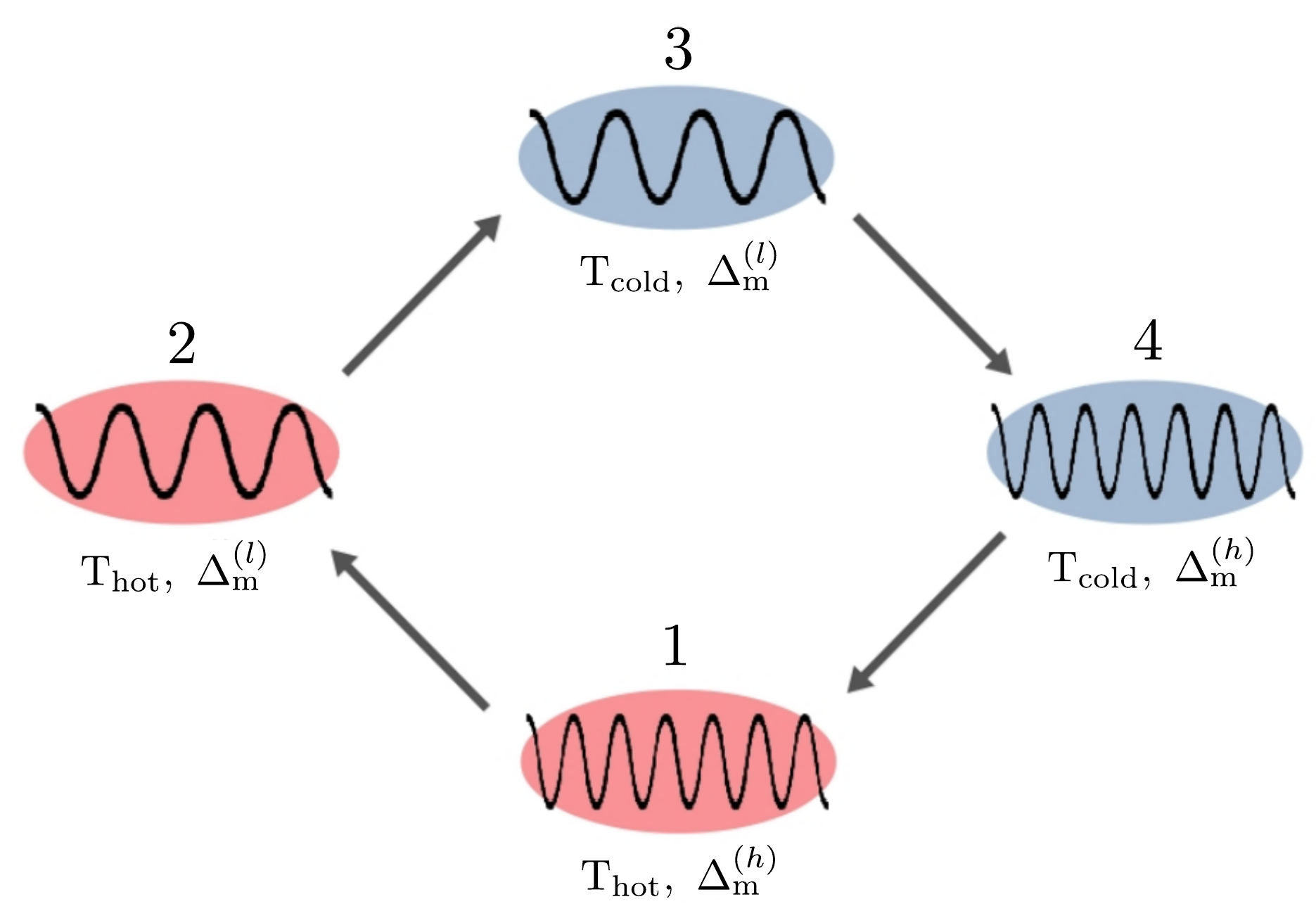}
\caption{The Stirling cycle: $1 \rightarrow 2$, isothermal stroke (expansion at constant temperature);
$2 \rightarrow 3$, isochoric stroke (heat removal at constant volume);
$3 \rightarrow 4$, isothermal stroke (compression at constant temperature);
$4 \rightarrow 1$, isochoric stroke (heat addition at constant volume).}
\label{Fig2}
\end{figure}
\begin{figure}[t!]
\centering
\includegraphics[width=\columnwidth]{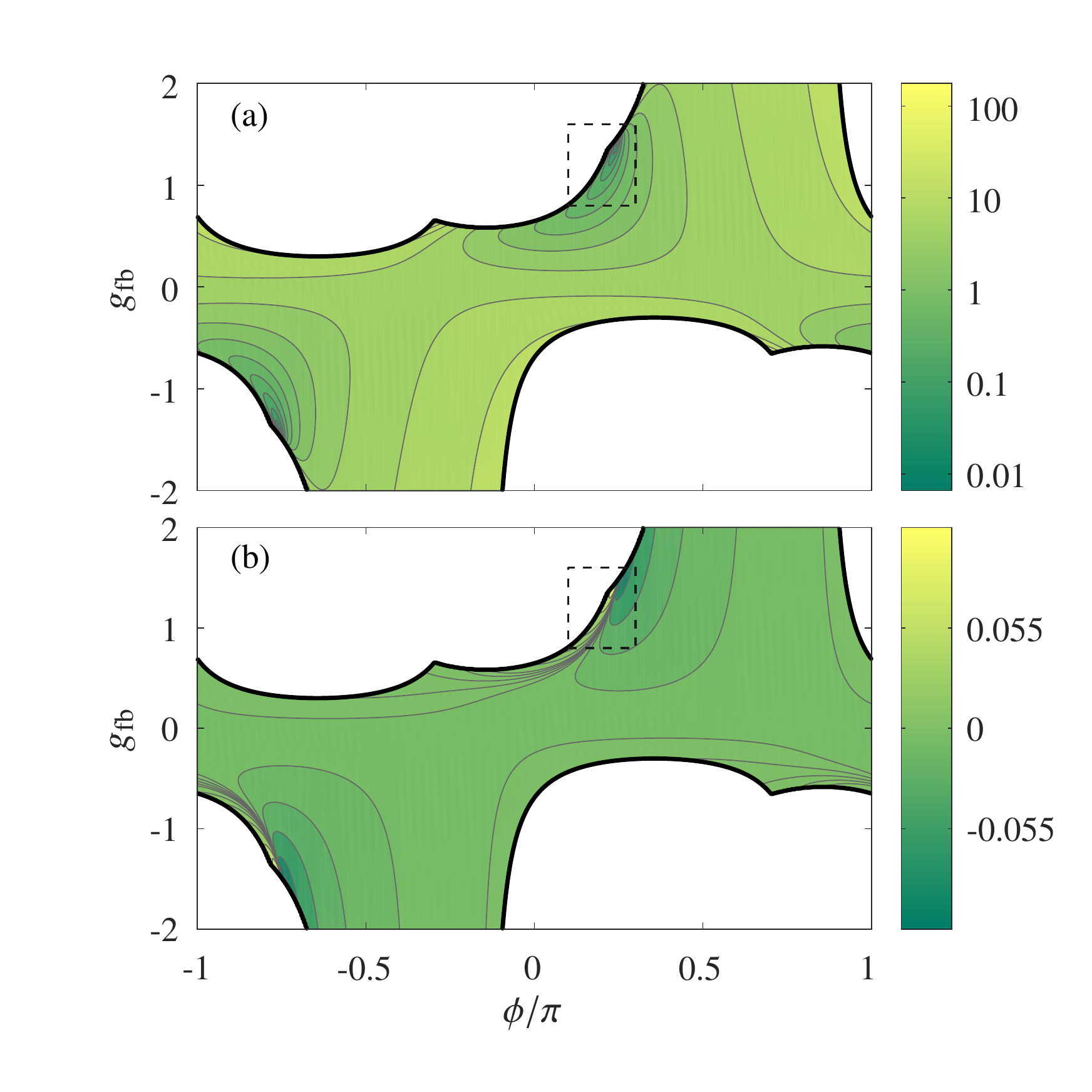}\\
\caption{Contour plot of (a) the effective temperature $T(\phi,g_\fb)$ (in kelvin) and (b) the optical spring $\Delta_\mm(\phi,g_\fb)$ (in units of $\omega_\mm$)
as a function of the feedback phase $\phi$ and gain $g_\fb$. The white areas indicate the parameter regime in which the system becomes unstable, 
i.e., in these areas the real part of at least one of the eigenvalues of the drift matrix of the system of quantum Langevin equations~\rp{dotaeff} and \rp{dotb} (and their Hermitian conjugates) is positive. 
The areas, enclosed by the dashed squares, are magnified in Fig.~\ref{Fig4}.
We assume a mechanical oscillator frequency $\omega_\mm=2\pi \times 100$~kHz 
operated at a room temperature of 300~K.
The other parameters are $G=0.1\, \omega_\mm$, $\kappa_1=\kappa_2= \omega_\mm$, $\Delta=\omega_\mm$, $\gamma=10^{-4}\omega_\mm$, 
and detection efficiency $\eta_d=0.9$. 
}
\label{Fig3}
\end{figure}
\begin{figure}[t]
\centering
\includegraphics[width=\columnwidth]{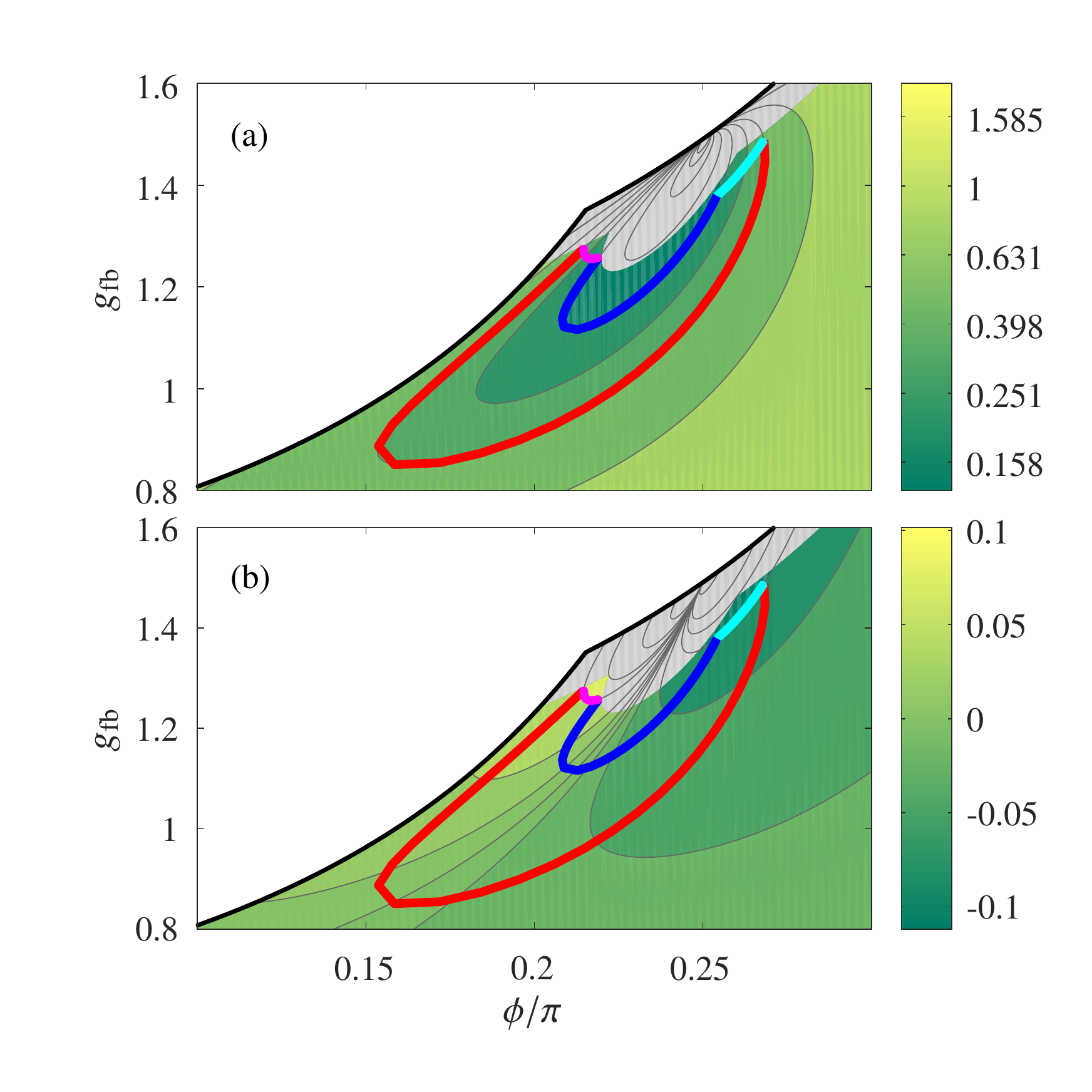}\\
\caption{Contour plot of (a) the effective temperature $T(\phi,g_\fb)$, in kelvin, and (b) the optical spring $\Delta_\mm(\phi,g_\fb)$
(in units of $\omega_\mm$) as a function of the feedback phase $\phi$ and gain $g_\fb$, corresponding to the dashed squares in Fig.~\ref{Fig3}. 
The white areas indicate the parameter range in which the system becomes unstable. 
The gray areas indicate the parameter values for which $\kappa_\eff\leq G,\Gamma_\mm$, where the adiabatic elimination is no longer valid.
The thick red and blue lines correspond to the isotherms (the red line at $T_{\rm hot}=0.5$~K and the blue one at $T_{\rm cold}=0.22$~K) of the engine cycle. 
The pink and cyan lines denote the isochores (with $\Delta_\mm\al{h}=0.08\, \omega_\mm$ for the pink line 
and $\Delta_\mm\al{l}=-0.08\, \omega_\mm$ for the cyan one). 
The other parameters are the same as in Fig.~\ref{Fig3}.}
\label{Fig4}
\end{figure}

\begin{figure}[t!]
\centering
\includegraphics[width=\columnwidth]{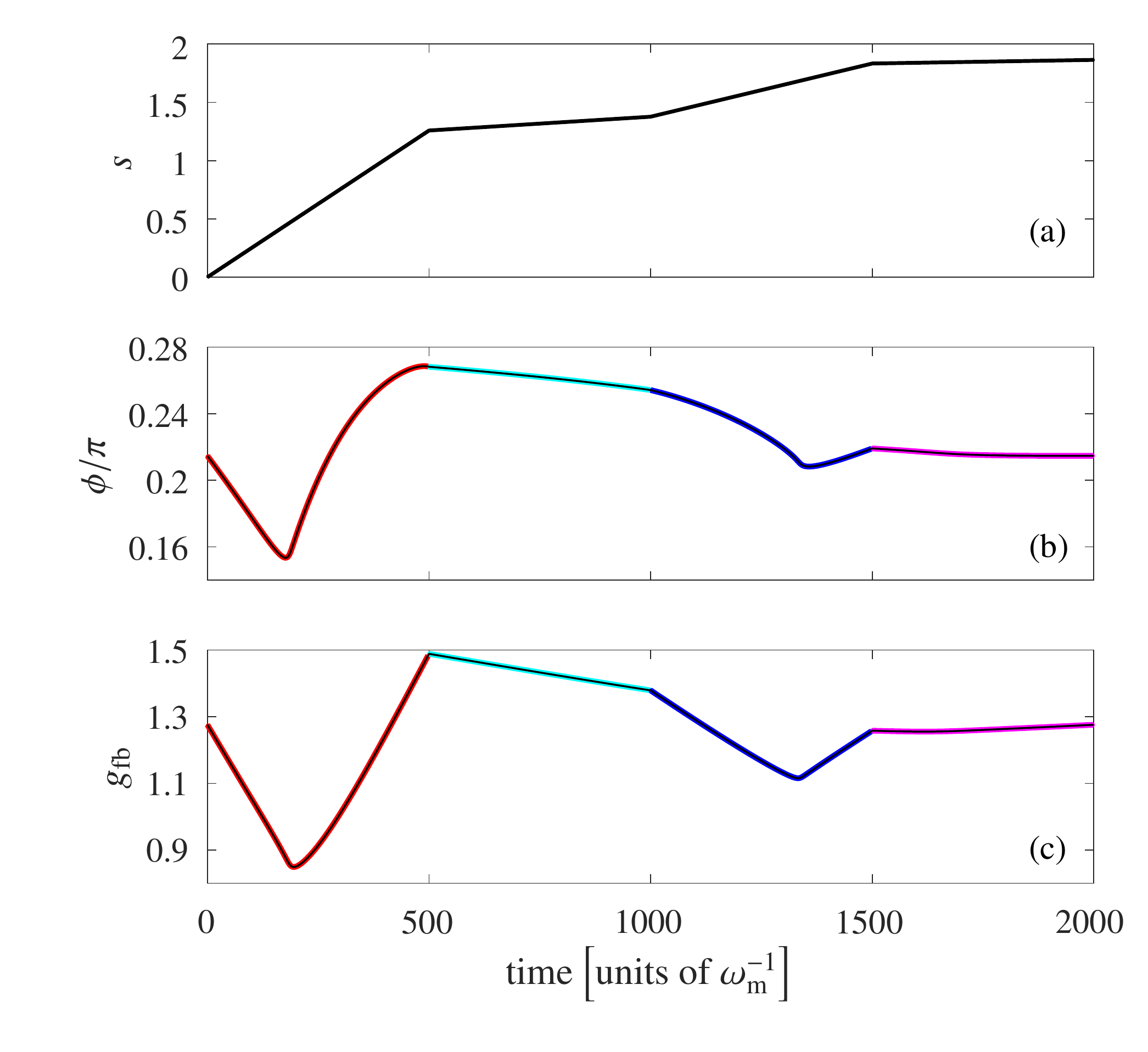}
\caption{Time evolution of the feedback parameters along the four strokes of the Stirling cycle depicted in Fig.~\ref{Fig4}. In (a), the quantity $s$ is the path length covered by the system along the cycle lines depicted in Fig.~\ref{Fig4}. The time variation of $s$ during each stroke is linear, i.e., the speed in each stroke is constant. The corresponding time changes of the feedback phase and gain are reported below in panels (b) and (c), respectively.}
\label{Fig5}
\end{figure}
\begin{figure}[h!]
\centering
\includegraphics[width=\columnwidth]{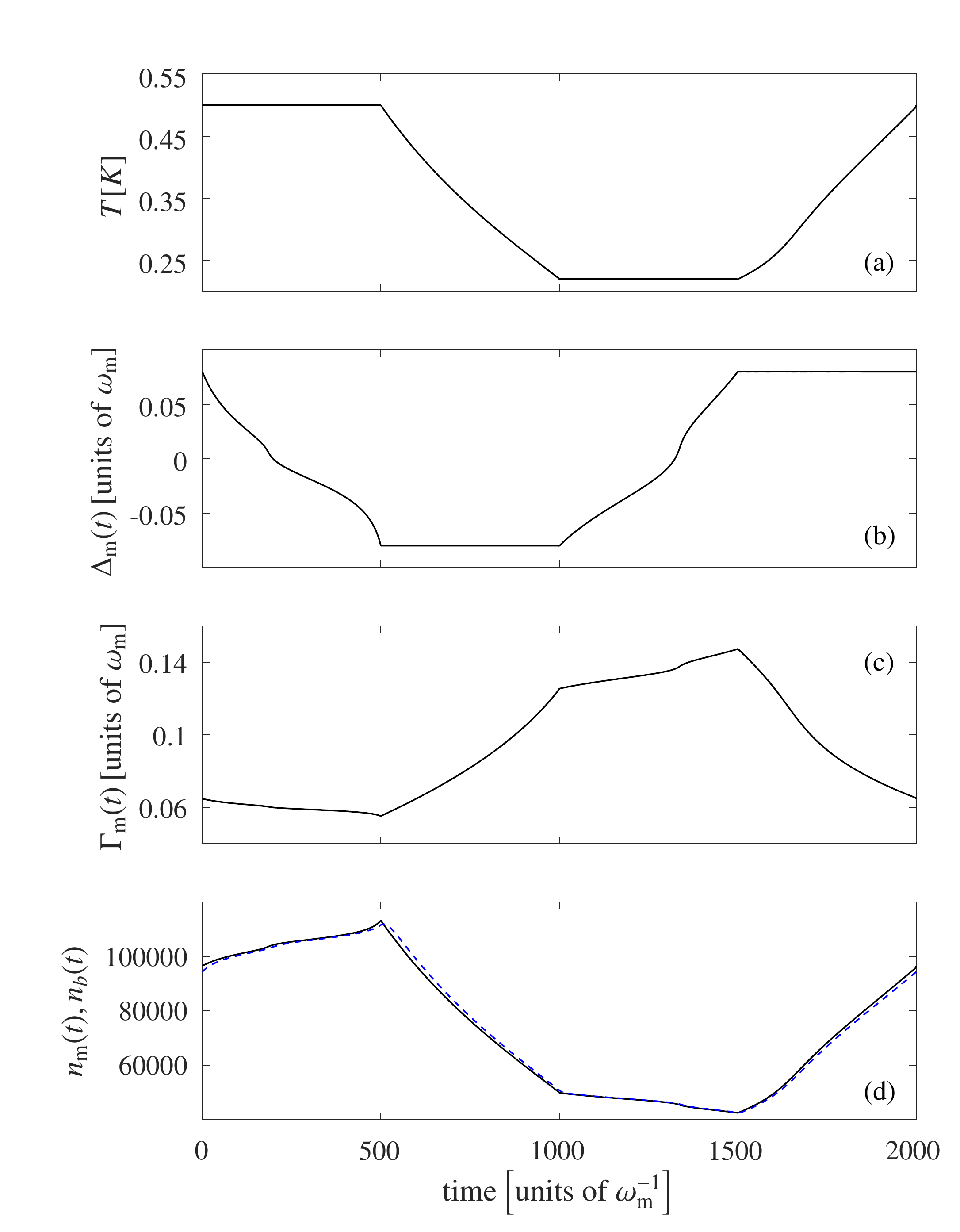}
\caption{Time evolution of the effective system parameters 
($T, \Delta_\mm$, $\Gamma_\mm$, $n_\mm$, and $n_b$)
along the four strokes of the Stirling cycle depicted in Fig.~\ref{Fig4}, and corresponding to the feedback parameters reported in Fig.~\ref{Fig5}.
In (d) the solid black line is for $n_b$ and the dashed blue one for $n_\mm$.
In this case the resulting engine efficiency is $\eta= 0.13$.}
\label{Fig6}
\end{figure}
\begin{figure}[h!]
\centering
\includegraphics[width=\columnwidth]{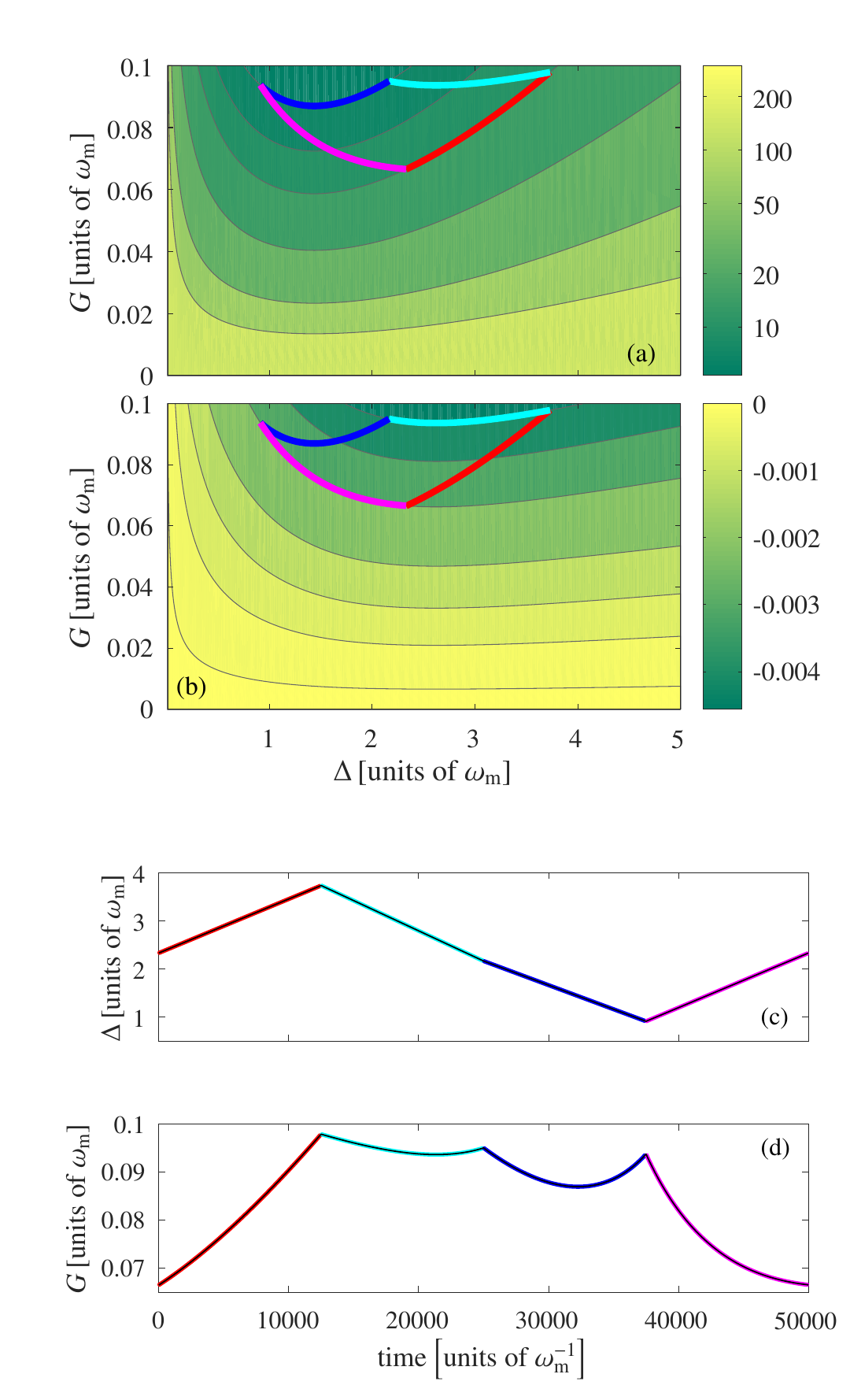}
\caption{Contour plot of (a) the effective temperature $T(\Delta,G)$ 
(in kelvin) and (b) the optical spring $\Delta_\mm(\Delta,G)$ 
(in units of $\omega_\mm$) without feedback ($g_\fb=0$), as a function of the laser detuning $\Delta$ and of the optomechanical interaction strength $G$. A possible Stirling cycle is also depicted: 
the thick red and blue lines correspond to the isotherms (the red at $T_{\rm hot}=15$~K and the blue at $T_{\rm cold}=7$~K) of the engine cycle. 
The pink and cyan lines are the isochores (with $\Delta_\mm\al{h}=-0.002\, \omega_\mm$ for the pink line and $\Delta_\mm\al{l}=-0.004\, \omega_\mm$ for the cyan one).
The plots (c) and  (d) show the corresponding time evolution of the laser detuning $\Delta$ and of the coupling
strength $G$. The other parameters are as in Fig.~\ref{Fig3}. 
In this case the resulting efficiency is $\eta= 0.002$.}
\label{Fig7}
\end{figure}
\begin{figure}[h!]
\centering
\includegraphics[width=\columnwidth]{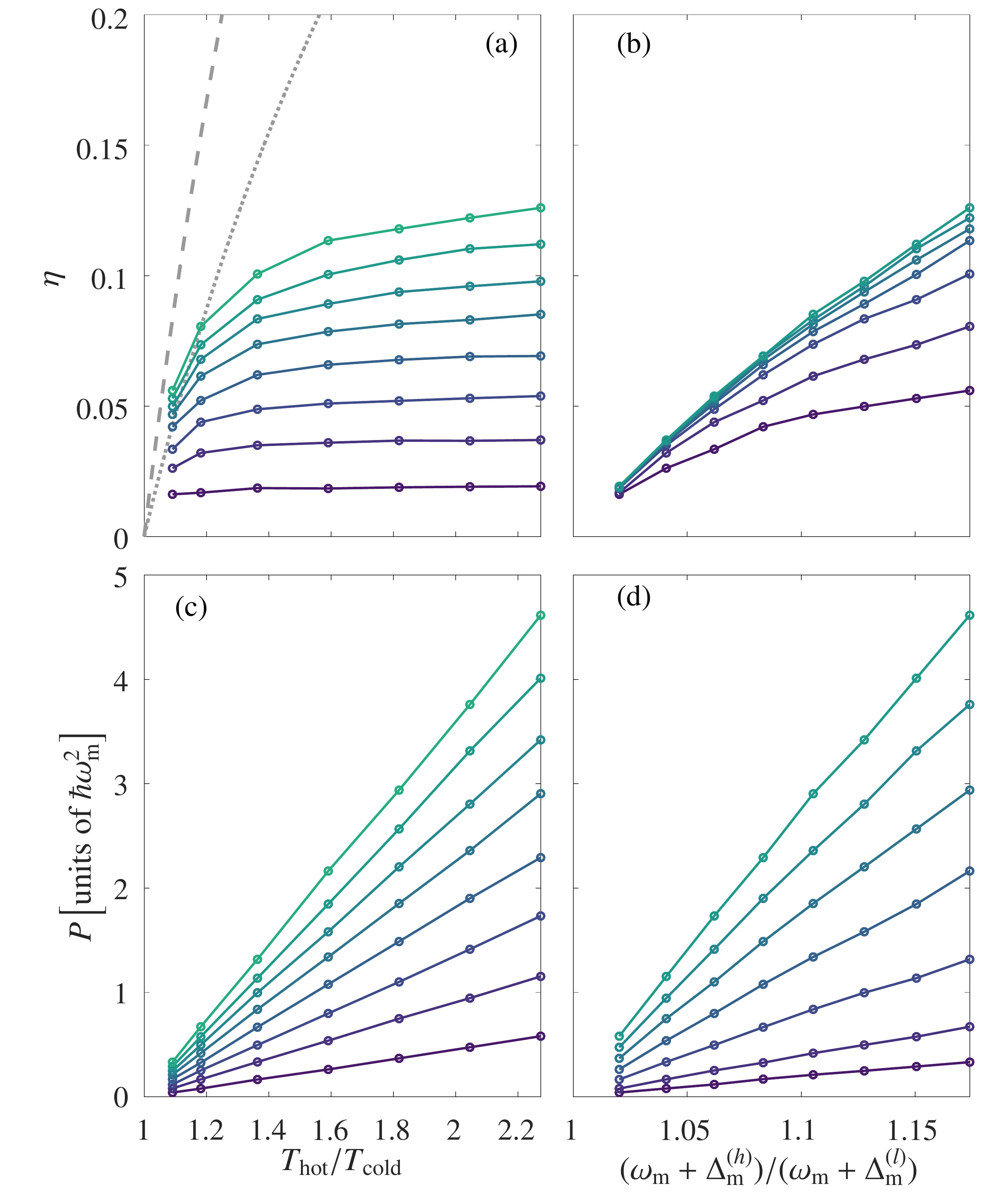}\\
\caption{Efficiency (upper panels) %(a)-(b) 
and power (lower panels) %(c)-(d) 
as a function of the ratio of: %(a)-(c) 
hot to cold temperature (left) and %(b)-(d) 
high to low mechanical frequency (right). 
In (a) the dashed gray line represents the Carnot efficiency $\eta_C \equiv 1-T_{\rm cold}/T_{\rm hot}$ and the dotted gray line the Curzon-Ahlborn 
efficiency $\eta_{C-A} \equiv 1-\sqrt{T_{\rm cold}/T_{\rm hot}}$. 
In (a) and (c) the lines from dark blue to light green are evaluated for 
equally spaced values of 
$\Delta_\mm\al{h}=-\Delta_\mm\al{l}$ 
in the range $[0.01,0.08]~\omega_\mm$.
In (b) and (d) the lines from dark blue to light green are evaluated for 
$T_{\rm cold}=0.22$~K, while $T_{\rm hot}$ takes the values $\{0.24,0.26,0.3,0.35,0.4,0.45,0.5\}$~K.
The other parameters are as in Fig.~\ref{Fig3}.}
\label{Fig8}
\end{figure}
\begin{figure}[h!]
\centering
\includegraphics[width=\columnwidth]{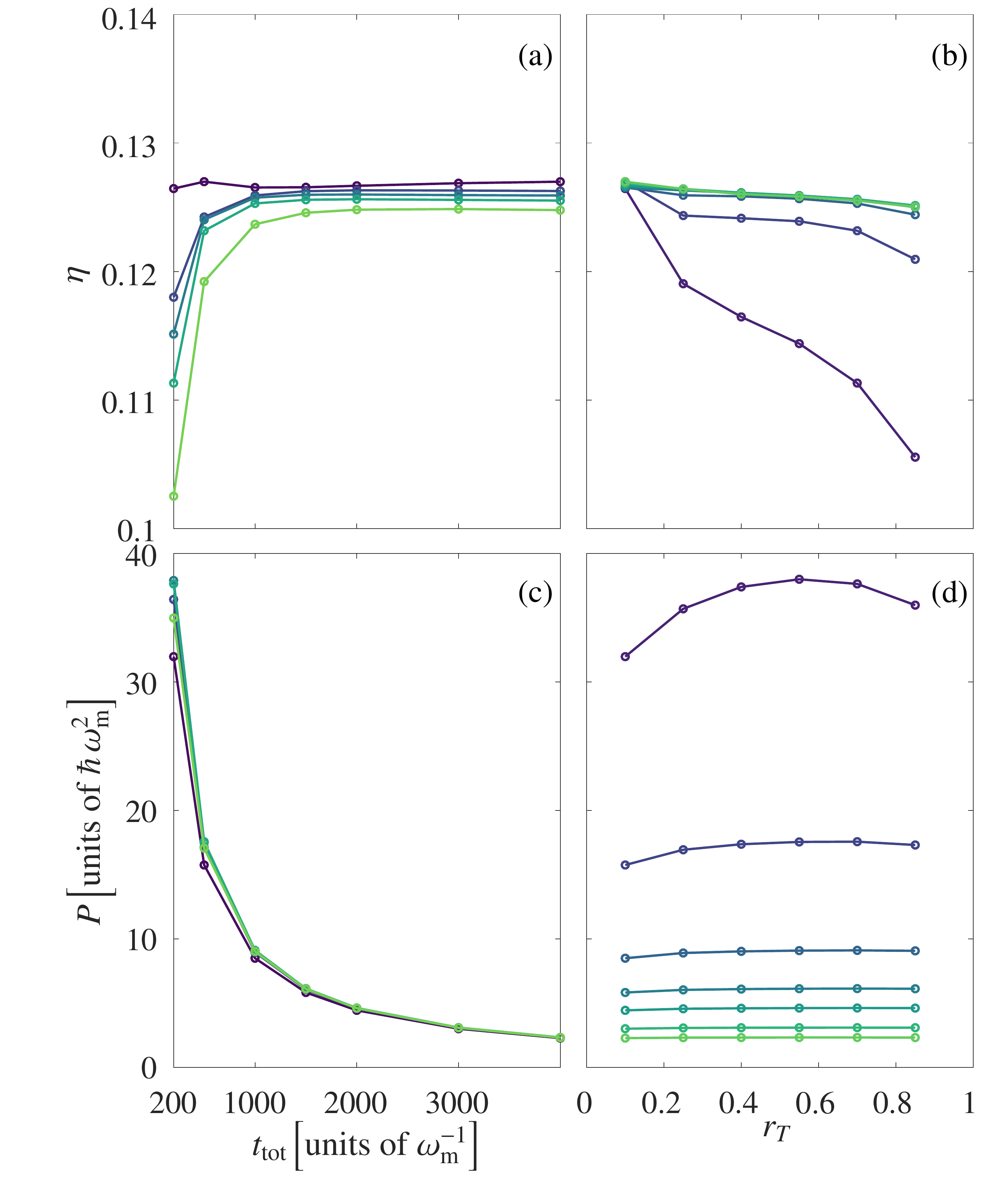}\\
\caption{Efficiency (upper panels) %(a)-(b) 
and power (lower panels) %(c)-(d) 
as a function of %(a)-(c) 
the total cycle time $t_{\rm tot}$ (left) and %(b)-(d) 
of the ratio of isothermal duration to the total time $r_T$ (right).
In (a) and (c) the lines from dark blue to light green are evaluated for 
equally spaced values of 
$r_T$ in the range $[0.1,0.9]$.
In (b) and (d) the lines from dark blue to light green are evaluated for $t_{\rm tot}$ 
which takes the values $\{0.2,0.5,1,1.5,2,3,4\}\times 10^3\,\omega_\mm^{-1}$.
The other parameters are as in Fig.~\ref{Fig3}. 
}
\label{Fig9}
\end{figure}
\begin{figure}[h!]
\centering
\includegraphics[width=\columnwidth]{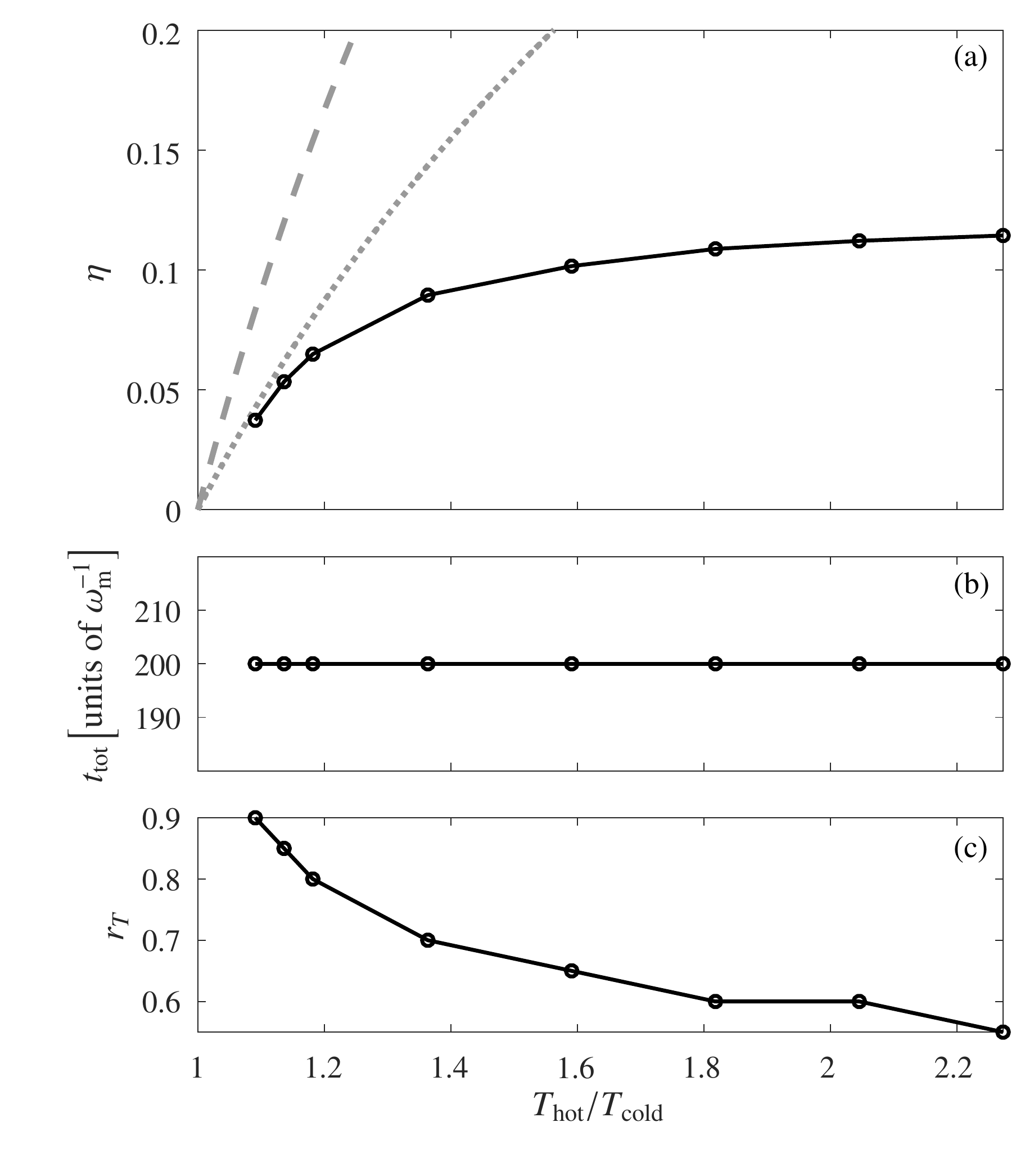}\\
\caption{(a) Efficiency at maximum power as a function of the ratio of hot to cold bath temperature, 
and corresponding values of (b) the total cycle time $t_{\rm tot}$ and (c) the ratio of isothermal duration to the total time $r_T$. 
In (a) the dashed gray line represents the Carnot efficiency, whereas the dotted gray line the Curzon-Ahlborn 
efficiency. 
The other parameters are as in Fig.~\ref{Fig3}.
}
\label{Fig10}
\end{figure}

\section{The optomechanical Stirling cycle}
\label{Engine}
The Stirling engine works between two isochoric and two isothermal  transformations, as schematically depicted in Fig.~\ref{Fig2}. 
In our system these transformations are realized by controlling the feedback gain $g_\fb$ and the feedback phase $\phi$, introduced, respectively, in Eqs.~\rp{ain1} and \rp{fb_phase}.
By adjusting these parameters, it is possible to tune the cavity response function, Eq.~\rp{Lambda}.
The latter, in turn, affects the optical spring $\Delta_\mm(\phi,g_\fb)$, Eq.~\rp{Deltam},
and the number of effective thermal excitations $n_\mm(\phi,g_\fb)$, Eq.~\rp{nm}, which, hence, both depend on $g_\fb$ and $\phi$.    
Correspondingly this allows to have control over the effective temperature of the reservoir, which is given by 
\begin{eqnarray}
T(\phi,g_\fb)=\frac{\hbar\pq{\omega_\mm+\Delta_\mm(\phi,g_\fb)}\,n_\mm(\phi,g_\fb)}{k_B}\,,
\end{eqnarray} 
with $k_B$ being Boltzmann's constant. 
So, despite the system is operated at room temperature, the applied laser cooling allows to engineer
effective bath temperatures below $1$~K. 
Therefore, by tuning the feedback parameters it is possible to steer the dynamics of the mechanical resonator and to drive it through specific thermodynamic transformations.

The variation of the feedback parameters should take place on a sufficiently slow time scale, such that the approximations introduced in the previous section remain valid. 
In particular, the adiabatic elimination is valid as long as the cavity field is always well approximated by its instantaneous steady state.
In this case, it is possible to tune the feedback and steer the mechanical oscillator through 
a Stirling cycle (see Fig.~\ref{Fig2}). In analogy to the work done by a piston, here the work corresponds to energy variations due to changes in a Hamiltonian parameter (that is, in the present case, the mechanical frequency) which, therefore, plays the role of state variable such as the volume of conventional thermodynamical heat engines~\cite{zhang2014a,dechant2015a,klaers2017}.
The first stroke ($1 \rightarrow 2$) of Fig.~\ref{Fig2} realizes
the first isothermal expansion, 
where the vibrational frequency decreases from the initial 
upper value $\omega_\mm+\Delta_\mm\al{h}$ to the final lower value $\omega_\mm+\Delta_\mm\al{l}$.
In the second stroke, corresponding to the isochoric transformation $2 \rightarrow 3$, heat is removed from the system and the temperature is lowered from $T_{\rm hot}$ to $T_{\rm cold}$. 
The second isothermal corresponds to the compression stage (third stroke $3 \rightarrow 4$), where the mechanical frequency is increased back to its initial value. 
Finally, in the last stroke $4 \rightarrow 1$, heat is absorbed at constant frequency and the temperature returns to its initial value.

In order to determine the values of $g_\fb$ and $\phi$, which allow to perform the Stirling cycle, we numerically explored the behaviour of the effective temperature [see Fig.~\ref{Fig3} (a)]
and of the optical spring [see Fig.~\ref{Fig3} (b)] as a function of these parameters. 
Level lines in these two contour plots represent isotherms and isochores, respectively. 
Our aim is to identify a pair of isotherms, at different temperatures, and a pair of isochores, at different mechanical frequencies, which form a closed loop, i.e., 
a Stirling cycle, in the $(\phi, g_\fb)$ plane.

Figure~\ref{Fig3} shows that maximum variability of the optomechanical parameters is achieved close to the system instability (represented by the white areas). 
Hence, we confine ourselves to the region enclosed by the dashed square, in Fig.~\ref{Fig3}, and identify a few specific isothermal and isochoric lines, suitable to implement a Stirling cycle. 
The magnified view of this region is depicted in Fig.~\ref{Fig4}. In particular, we focus on the area where the effective cavity decay rate is $\kappa_\eff> G,\Gamma_\mm$, such that the cavity dynamics is fast enough to assure the validity of the adiabatic elimination.
Figure~\ref{Fig4} highlights a specific Stirling cycle where two isotherms (blue and red lines) cross two isochores (cyan and pink lines) in order to form a closed loop.
In particular, the blue line denotes the cold isotherm, the red line corresponds to the hot one, the cyan line marks the low frequency isochore, and the pink one is the high frequency one. 

In order to simulate the mechanical oscillator dynamics during a Stirling cycle,  
the values of the feedback parameters $\phi$ and $g_\fb$ are properly varied so that, in each stroke, the system moves along the path in the $\pt{\phi,g_\fb}$ space 
at constant speed (see Fig.~\ref{Fig5}). 
This results in a nonlinear time evolution of $\phi(t)$ and $g_\fb(t)$, as depicted in Figs.~\ref{Fig5} (b) and (c). 
The values of $\phi(t)$ and $g_\fb(t)$ determine the time evolution of the other system parameters $\Gamma_\mm(t)$, $\Delta_\mm(t)$, and $n_\mm(t)$ (see Fig.~\ref{Fig6}), 
which appear in the reduced equations of the mechanical resonator Eqs.~\rp{dotb_adel} and \rp{ndot}.
The isothermal strokes are characterized by a constant effective temperature, determined by the feedback parameters [Fig.~\ref{Fig6}(a)].
Instead, the optical spring $\Delta_\mm(t)$ remains constant during the isochoric transformations [Fig.~\ref{Fig6}(b)].
The corresponding dynamics of the effective damping rate $\Gamma_\mm$, the number of system excitations $n_b$, and the effective bath excitations $n_\mm$ are shown
in Fig.~\ref{Fig6}(c) and (d).
The behaviour of $n_b$ is evaluated by numerically solving Eq.~\rp{ndot}, with the time dependent parameters reported in Fig.~\ref{Fig6} and using the instantaneous steady state as initial condition.
In order to analyze the proper working regime of the engine, we have computed the time evolution of the system over several cycles and verified that, after a few cycles, the resonator dynamics stabilizes and repeats itself from cycle to cycle.

In this work, we have considered only linear variations of the system along the cycle lines in the $\pt{\phi,g_\fb}$ space. However, the optimization of the engine performances would probably require a more sophisticated and customized control of the feedback parameters, similar to the approach discussed in Ref.~\cite{dechant2015a} or by using techniques such as shortcut-to-adiabaticity~\cite{abah2017}.

\section{Engine efficiency and power}
\label{Efficiency}
In order to assess the engine performance, we have evaluated its efficiency and power delivery. 
The efficiency is defined as the ratio of the work $W_{\rm tot}$ done by the engine to the absorbed heat $Q_{\rm abs}$
\begin{eqnarray}
\eta \equiv \frac{-W_{\rm tot}}{Q_{\rm abs}}\ .
\end{eqnarray}
The power, instead, is by definition the work done per unit time
\begin{eqnarray}
P \equiv \frac{-W_{\rm tot}}{t_{\rm tot}}\ ,
\end{eqnarray}
where $t_{\rm tot}$ is the cycle duration.
In our notation we consider positive quantities both the heat $Q$, absorbed by the system, and the work $W$, performed by the environment 
on the system. Therefore, the work done by the engine is negative and the variation of the internal energy is $\Delta U=Q+W$.
For a quantum system the internal energy $U$ is given by the expectation value of the Hamiltonian,
$U(t)=\av{H(t)}$.
In the present case the 
Hamiltonian 
for the mechanical resonator, after the adiabatic elimination of the cavity field (see Eq.~\rp{dotb_adel}),
is
\begin{eqnarray}
H(t)&=&\hbar\pq{\omega_\mm +\Delta_\mm(t)}b\da(t) b(t)\ ,
\end{eqnarray}
and, thus, the internal energy is given by
\begin{eqnarray}\label{Ut}
U(t)&=&
\hbar\pq{\omega_\mm +\Delta_\mm(t)}\,n_b(t)
\ ,
\end{eqnarray}
where $n_b(t)$ is the average number of mechanical excitations described by Eq.~\rp{ndot}.
The heat exchanged from the initial time $t_i$ to the final time $t_f$ is given by the integral~\cite{abari2019} 
\begin{eqnarray}
Q=\int_{t_i}^{t_f}\dd t\ {\rm Tr}\pg{H(t)\ \dot\rho(t)}\ ,
\end{eqnarray}
where, here, $\rho(t)$ is the density matrix for the mechanical resonator. It fulfills
the Lindblad master equation, which is equivalent to the quantum Langevin equation~\rp{dotb_adel},
\begin{eqnarray}
\dot\rho(t)&=&-\ii\pq{H(t),\rho(t)} + \frac{\gamma+\Gamma_\mm(t)}{2} \lpg{
\pq{n_\mm(t)+1}\, \LL[b]\cdot\rho(t)^{\phantom{\dagger}} 
}\nn\\&&\rpg{
+\ n_\mm(t)\ \LL[b\da]\cdot\rho(t)
}\ ,
\end{eqnarray}	
with $\LL[x]\cdot\rho=2\,x\,\rho\,x\da-x\da x\,\rho-\rho\,x\da x$. Thereby one finds
\begin{eqnarray}\label{Q}
Q=\hbar\int_{t_i}^{t_f}\dd t\ \pq{\gamma+\Gamma_\mm(t)}\pq{\omega_\mm+\Delta_\mm(t)}\pq{n_\mm(t)-n_b(t)} .
\end{eqnarray}
Eventually one can determine the work as the difference between the variation of the internal energy and the heat.

We have computed  the heat and work exchanged by the system by solving numerically Eq.~\rp{ndot}, with the time dependent coefficients that follow various thermodynamic cycles, similar to the ones reported in Fig.~\ref{Fig4}, and we have analyzed the corresponding engine performance in terms of efficiency and power. 
For instance, in the case of the cycle of Fig.~\ref{Fig4}, the resulting efficiency is $\eta=0.13$. \\
We have to emphasize that, in principle, a similar engine could be accomplished even without feedback by adjusting the temperature and the optical spring via the laser intensity, which determines the optomechanical interaction strength $G$, and the laser detuning $\Delta$. 
However, as shown in 
Fig.~\ref{Fig7}, 
this approach would produce a much smaller tunability of the optical spring, which in turn would result in very low engine efficiencies (in the case of Fig.~\ref{Fig7}, which is obtained with parameters consistent with those used in Fig.~\ref{Fig4}, $\eta=0.002$).
The feedback, instead, allows to achieve a sufficiently large variation of the optical spring and, consequently, sizeable efficiencies.

Figure~\ref{Fig8} shows the calculated efficiency and power for different values of temperature and mechanical frequency. In particular, we explore how the performance of the engine depends on the ratio between the hot and cold isotherm temperature, $T_{\rm hot}/T_{\rm cold}$, and
on the compression ratio, which, in our case, corresponds to the ratio between the high and low mechanical frequencies $(\omega_\mm+\Delta_\mm\al{h})/(\omega_\mm+\Delta_\mm\al{l})$.
Both efficiency and power increase with increasing values of these two quantities.
As expected the efficiency remains always below the classical limit set by the Carnot efficiency 
\begin{eqnarray}
\eta_C \equiv 1-\frac{T_{\rm cold}}{T_{\rm hot}}
\end{eqnarray}
of an ideal engine operating between the same two isotherms [see the dashed gray line in Fig.~\ref{Fig8} (a)]. 
Moreover, we note that it is not possible to increase $T_{\rm hot}/T_{\rm cold}$ or $(\omega_\mm+\Delta_\mm\al{h})/(\omega_\mm+\Delta_\mm\al{l})$ beyond certain values.
For example, in the case of Fig.~\ref{Fig4}, increasing either $T_{\rm hot}/T_{\rm cold}$ or $(\omega_\mm+\Delta_\mm\al{h})/(\omega_\mm+\Delta_\mm\al{l})$ would bring the system in the regime in which $\kappa_\eff<G$, where the adiabatic elimination is no longer valid. 
This implies that the results of Figs.~\ref{Fig4}-\ref{Fig6} cannot be further improved by selecting different temperatures or mechanical frequencies.

The previous results have been achieved considering a fixed total cycle time and an equal duration for each stroke.
In Fig.~\ref{Fig9} we show how the performance of the engine depend on the duration of the cycle. 
In particular, we present the results as a function of the total time $t_{\rm tot}$ [see Fig.~\ref{Fig9} (a) and (c)]  and of the ratio $r_T$ between the duration of the isothermal strokes and the total time (assuming, however, that both the two isotherms and the two isochores have equal duration) [see Fig.~\ref{Fig9} (b) and (d)].
We observe that the dependence of the efficiency on the cycle length is relatively weak. 
Instead, the power rapidly decreases with $t_{\rm tot}$, while it is not too affected by changes of $r_T$.
In Fig.~\ref{Fig10} (a), we report the corresponding efficiency at maximum power, as a function of the temperature ratio. Namely, we plot the efficiency evaluated at the specific values of $t_{\rm tot}$ and $r_T$ [reported in Figs.~\ref{Fig10} (b) and (c)], which result in the maximum power for each value of the temperature ratio.
As we have seen in Fig.~\ref{Fig9} the maximum power is always achieved for the smallest total time $t_{\rm tot}=200\,\omega_\mm^{-1}$.
As expected the efficiency is upper-bounded  by the Curzon-Ahlborn efficiency
\begin{eqnarray}
\eta_{C-A} \equiv 1-\sqrt{\frac{T_{\rm cold}}{T_{\rm hot}}}\ .
\end{eqnarray}

A final comment is in order. The scheme that we have analized is effective if the optomechanical coupling strength $G$ is smaller then the cavity decay rate $\kappa$ (see App.~\ref{App-rsl}). In fact the feedback permits  to effectively modify the cavity linewidth, see Eq.~\rp{kappaeff-Deltaeff}, and, in turn, this allows to have control over the mechanical parameters. However, if $G$ approaches the value of $\kappa$, the range of the feedback parameters, which are consistent with our analysis, shrinks
(because the adiabatic elimination is valid if $\kappa_\eff>G$, so that the value of $\kappa_\eff$ cannot be smaller than that of $\kappa$) and, as a consequence, the ability to design suitable thermodynamic cycles is reduced. 
In App.~\ref{App-rsl} we have reported an example (see Figs.~\ref{Fig13} and \ref{Fig14}), which describes this situation and shows that when $G=\kappa$, the efficiencies of the engine with and without feedback are comparable.

\section{Conclusions}
\label{Conclusions}

We have shown how to steer a mechanical resonator through a thermodynamic cycle using an in-loop cavity.
The cavity light is controlled by a feedback loop and is used both to engineer the reservoir by laser cooling, and to tune the mechanical frequency via the optical spring effect.  
Thereby, the engine cycle is achieved by controlling only the feedback transfer function,
which defines the feedback gain and the phase. This makes the experimental implementation of the proposed setup relatively simple.

The feedback system that we have analyzed allows to increase significantly the range of variability of the mechanical temperature (see Ref.~\cite{rossi2017}) and frequency, as compared to what is achievable by controlling the driving light frequency and intensity. 
Hence, we have shown that this effect can be employed to enhance the efficiency of a Stirling heat engine
over the efficiency achievable in a similar system without feedback 
by almost two orders of magnitude. 
We note that it might be worth considering if higher efficiencies could be attained by combining the two techniques, for example, by controlling the driving laser power to engineer the reservoir number of excitations and using the feedback for tuning the mechanical frequency.
Moreover, a further increase of the engine efficiency could be achieved by resorting to a more elaborate time control of the system parameters~\cite{gomez-marin2008,dechant2015a} and by implementing shortcut-to-adiabaticity techniques~\cite{abah2017}.
In order to analyze heat and work experimentally, one should determine the time evolution of temperature, mechanical frequency, and mechanical dissipation rate during the engine cycle and, then, use the formulas~\rp{Ut} and \rp{Q} to compute the corresponding work and heat. These quantities can be determined by probing the system with a light field resonant with a different
optical cavity mode and measuring the phase modulation of the transmitted 
or reflected field. The corresponding power spectrum is proportional to the position spectrum of the mechanical mode, from which it is possible to extract the mechanical variables (see, for example, Ref.~\cite{rossi2017}). In order to resolve the mechanical peak, the spectrum should be evaluated over a sufficiently long time interval, 
%that should be 
much longer than the inverse of the mechanical linewidth (the dissipation rate). As shown in Fig.~\ref{Fig6}, the cycle time is so long that it should be possible to faithfully resolve the position spectrum at different times and, therefore, to reconstruct the time evolution of the mechanical variables.
  
Our analysis is performed in a regime in which quantum phenomena are not yet observable. 
In fact, here we are interested in indicating the simplest route towards a first experimental realization of an optomechanical heat engine. Nevertheless, it would be very interesting to study in detail regimes where quantum effects are relevant and novel quantum thermodynamical processes could be explored. In order to move towards this direction it would be necessary to relax also other assumptions that are at the basis of the present investigation.
For example, we have analyzed only situations in which the system-reservoir coupling (i.e., the optomechanical coupling) is weak. 
Moreover, our results have been obtained for relatively slow transformations, such that the evolution of the number of mechanical excitations of the resonator closely follows that of the bath [see Fig.~\ref{Fig6}(d)].
If either the coupling is not sufficiently small or the transformations are not sufficiently slow, the adiabatic elimination, at the basis of our investigation, is no more valid. In these cases our study should be extended by including a detailed analysis of the fully coupled optomechanical dynamics.
In this regime it would be particularly interesting to investigate the role of correlations between the system and the reservoir~\cite{gelbwaser-klimovsky2015b,uzdin2016,wiedmann2020a,newman2020},
to study effects related to correlations in the reservoir~\cite{abah2014a,niedenzu2016} (which in this system can be generated by the feedback itself~\cite{zippilli2018}), to analyze if this feedback setup can play the role of a Maxwell demon~\cite{ribezzi-crivellari2019},  and, more generally, to examine the role of the feedback in the energy exchanges of quantum heat engines~\cite{campisi2017,potts2018}.

%%%%%%%%%%%%%%%%%%%%%%%%%%%%%%%%%%%%%%%%%%%%%%%%%%%%%%%%%%%%%%%%%%%%%%

\appendix

\section{Adiabatic elimination of the cavity field}
\label{App-ad-el}

In this appendix we  discuss the derivation of the effective equations for the mechanical resonator~Eqs.~\rp{dotb_adel} and \rp{ndot}, which are valid in the weak coupling regime $G\lesssim\kappa_\eff$ and are determined by adiabatically eliminating the cavity field from the Langevin equations of the mechanical resonator. 
Specifically we focus on the slowly varying mechanical operator $\bar b(t)$ defined by the transformation $b(t) \equiv \ee^{-\ii\,\omega_\mm t}\ \bar b(t)$, 
in comparison to which the cavity field evolves on a much shorter time scale. 
Hence, we first determine the steady state for the cavity field and then substitute it in the equation for $\bar b(t)$. 
The steady state for the cavity field operator can be expressed in terms of its Fourier transform with $\tilde x(\omega)\equiv\frac{1}{\sqrt{2\pi}}\int\,\dd t\ \ee^{\ii\,\omega \,t}x(t)$ and $\pq{\tilde x(\omega)}\da\equiv\tilde x\da(-\omega)$], introducing the 
cavity  response function modified by the feedback~\cite{zippilli2018} 
\begin{eqnarray}\label{lambdaApp}
\tilde\lambda(\omega)=\frac{\tilde\chi(\omega)\pq{1-\mu^*\,\tilde\chi(-\omega)^*}}{1-\abs{\mu}^2\,\tilde\chi(\omega)\,\tilde\chi(-\omega)^*}
\end{eqnarray}
with $\tilde\chi(\omega)$ defined in Eq.~\rp{chi}, as
\begin{eqnarray}\label{ast}
a(t)&=&a(t)\Bigr|_{G=0}-\frac{\ii\,G}{\sqrt{2\,\pi}}\int_{-\infty}^\infty\,\dd\omega\ \ee^{-\ii\,\omega\,t}\ 
\tilde\lambda(\omega)
\pq{\tilde b(\omega)+\tilde b\da(\omega)} \,,
\nn\\
\end{eqnarray}
where $a(t)\Bigr|_{G=0}=\frac{1}{\sqrt{2\,\pi}}\int_{-\infty}^\infty\,\dd\omega\ \ee^{-\ii\,\omega\,t}\ \tilde a(\omega)\Bigr|_{G=0}$ indicates the steady state field operator without resonator. The integral in Eq.~\rp{ast} can be further approximated, using the convolution theorem 
[$\frac{1}{\sqrt{2\,\pi}}\int\dd t\ \ee^{\ii\,\omega\,t}\int\dd t'\,f(t-t')g(t')=\sqrt{2\,\pi}\ \tilde f(\omega)\,\tilde g(\omega)$, 
i.e., $\int\dd t'\,f(t-t')g(t')=\int\dd \omega\ \ee^{-\ii\,\omega\,t}\,\tilde f(\omega)\,\tilde g(\omega)$] and introducing the slowly varying mechanical operators, 
which are essentially constant over the cavity time scale determined by the Fourier transform, $\lambda(t)$, of the response function $\tilde\lambda(\omega)$, as
\begin{eqnarray}
&&\int_{-\infty}^\infty\,\dd\omega\ \ee^{-\ii\,\omega\,t}\ \tilde\lambda(\omega)\ \tilde b(\omega)=\int_{-\infty}^\infty\,\dd t'\ \lambda(t-t')\ b(t')
\nn\\&&\hspace{1cm}
\simeq\bar b(t)\int_{-\infty}^\infty\,\dd t'\ \lambda(t-t')\ \ee^{-\ii\,\omega_\mm\,t}
\nn\\&&\hspace{1cm}
=\sqrt{2\,\pi}\ \bar b(t)\int_{-\infty}^\infty\,\dd\omega\ \ee^{-\ii\,\omega\,t}\ \tilde\lambda(\omega)\ \delta(\omega-\omega_\mm)
\nn\\&&\hspace{1cm}
=\sqrt{2\,\pi}\ \bar b(t)\ \ee^{-\ii\,\omega_\mm\,t}\ \tilde\lambda(\omega_\mm)\ ,
\end{eqnarray}
and similarly
\begin{equation}
\int_{-\infty}^\infty\,\dd\omega\ \ee^{-\ii\,\omega\,t}\ \tilde\lambda(\omega)\ \tilde b\da(\omega)
\simeq \sqrt{2\,\pi}\ \bar b\da(t)\ \ee^{\ii\,\omega_\mm\,t}\ \tilde\lambda(-\omega_\mm) \,.
\end{equation}
Thereby one finds
\begin{eqnarray}\label{ast0}
a(t)&\simeq&a(t)\Bigr|_{G=0}-\ii\,G\,\pg{
\tilde\lambda(\omega_\mm)
\,\ee^{-\ii\,\omega_\mm\,t}\ \bar b(t)
+\tilde\lambda(-\omega_\mm)
\,\ee^{\ii\,\omega_\mm\,t}\ \bar b\da(t)}\ ,
\nn\\
\end{eqnarray}
so that substituting it and its hermitian conjugate into the equation for $\bar b(t)$  and retaining only resonant terms  one finds Eq.~\rp{dotb_adel} of the main text,
where the modified noise operator $B_\inn(t)$ is explicitly given by
\begin{eqnarray}
B_\inn(t)=\frac{-\ii\,G\pq{a(t)\Bigr|_{G=0}+a\da(t)\Bigr|_{G=0}}+\sqrt{\gamma}\,b_\inn(t)}{\sqrt{\Gamma_\mm}}\,\ee^{\ii\,\omega_\mm\,t}\ ,
\end{eqnarray}
whose correlation functions can be expressed in terms of the correlations of the steady state cavity field quadrature 
\begin{eqnarray}\label{X0}
X_0(t) \equiv a(t)\Bigr|_{G=0}+a\da(t)\Bigr|_{G=0}
\end{eqnarray}
without resonator,
and of the mechanical input noise $b_\inn(t)$. 
In particular, according to our assumptions $\kappa>G,\gamma$
the correlation of $X_0(t)$ decays on a time scale smaller than the time scale of the mechanical dynamics.
Thus if we indicate with 
$o(t)$ a generic variable of the slow mechanical dynamics then we can approximate 
$\int\,\dd \tau\av{X_0(t)\ X_0(t+\tau)}\ o(\tau)
\simeq o(t)\int\,\dd \tau\av{X_0(t)\ X_0(t+\tau)}
=\frac{o(t)}{2\,\pi}\int\,\dd \tau\int\,\dd \omega\int\,\dd \omega'\,\ee^{-\ii\,\omega\,t}\ee^{-\ii\,\omega'\,(t+\tau)}\av{\tilde X_0(\omega)\ \tilde X_0(\omega')}
=\frac{o(t)}{2\,\pi}\int\,\dd \tau\int\,\dd \omega\,\ee^{\ii\,\omega\,\tau}S_{X_0}(\omega)
=o(t)\int\,\dd \omega\,\delta(\omega)\,\,S_{X_0}(\omega)
=o(t)\ S_{X_0}(0)
$ 
where we have introduced the power spectral density of the cavity field defined by the relation
\begin{eqnarray}\label{XXS}
\av{\tilde X_0(\omega)\ \tilde X_0(\omega')}=\delta(\omega+\omega')\ S_{X_0}(\omega)\ ,
\end{eqnarray}
and the specific form of which is reported below. 
This result implies that we can approximate the field quadrature correlation function as
\begin{equation}
\av{X_0(t)\ X_0(t+\tau)}\simeq\ \delta(\tau)\ S_{X_0}(0) \,.
\end{equation}
A similar calculation, including generic time-dependent phase factors, shows that 
\begin{eqnarray}
\av{X_0(t)\ X_0(t+\tau)}\ee^{\ii\,p\,t}\ee^{\ii\,q\,\tau}\simeq\ \delta(\tau)\ S_{X_0}(-q)\ \ee^{\ii\,p\,t}\ .
\end{eqnarray}
This approximation can be employed to determine the expressions for the correlation functions of the noise operator $B_\inn(t)$ reported in Eq.~\rp{BBcorrelations}.
The correlations $\av{B_\inn(t)\ B_\inn(t')}$ and $\av{B_\inn\da(t)\ B_\inn\da(t')}$, instead, include fast oscillating phase factors at frequency $\pm2\,\omega_\mm$.
Hence, their effect on the dynamics of the mechanical resonator is negligible.
Correspondingly, the equation for the mechanical excitation number is
\begin{equation}
\pd{t}\av{\bar b\da(t)\ \bar b(t)}=-\pt{\gamma+\Gamma_\mm}\ \av{\bar b\da(t)\ \bar b(t)}+G^2\ S_{X_0}(-\omega_\mm)+\gamma\,n_\T
\end{equation}
and similarly 
\begin{equation}
\pd{t}\av{\bar b(t)\ \bar b\da(t)}=-\pt{\gamma+\Gamma_\mm}\av{\bar b(t)\ \bar b\da(t)}+G^2\ S_{X_0}(\omega_\mm)+\gamma\,(n_\T+1)\,.
\end{equation}
Now, given that for bosonic operators  $\av{\bar b(t)\ \bar b\da(t)}-\av{\bar b\da(t)\ \bar b(t)}=1$, one finds that 
\begin{eqnarray}
\Gamma_\mm=G^2\,\pq{ S_{X_0}(\omega_\mm)-S_{X_0}(-\omega_\mm)}\ .
\end{eqnarray}
Thus, we finally find Eq.~\rp{ndot}.

Note that the response function defined in Eq.~\rp{lambdaApp} is the function which enters the equation for the field operator~\rp{ast}, while the one introduced in Eq.~(\ref{Lambda}) 
is the response function for the quadrature operator $X_0(t)$, see Eq.~\rp{tildeX0App}. They are related by the equation $\tilde\lambda(\omega)-\tilde\lambda(-\omega)^*=\Lambda(\omega)-\Lambda(-\omega)^*$.

\begin{figure}[t!]
\centering
\includegraphics[width=\columnwidth]{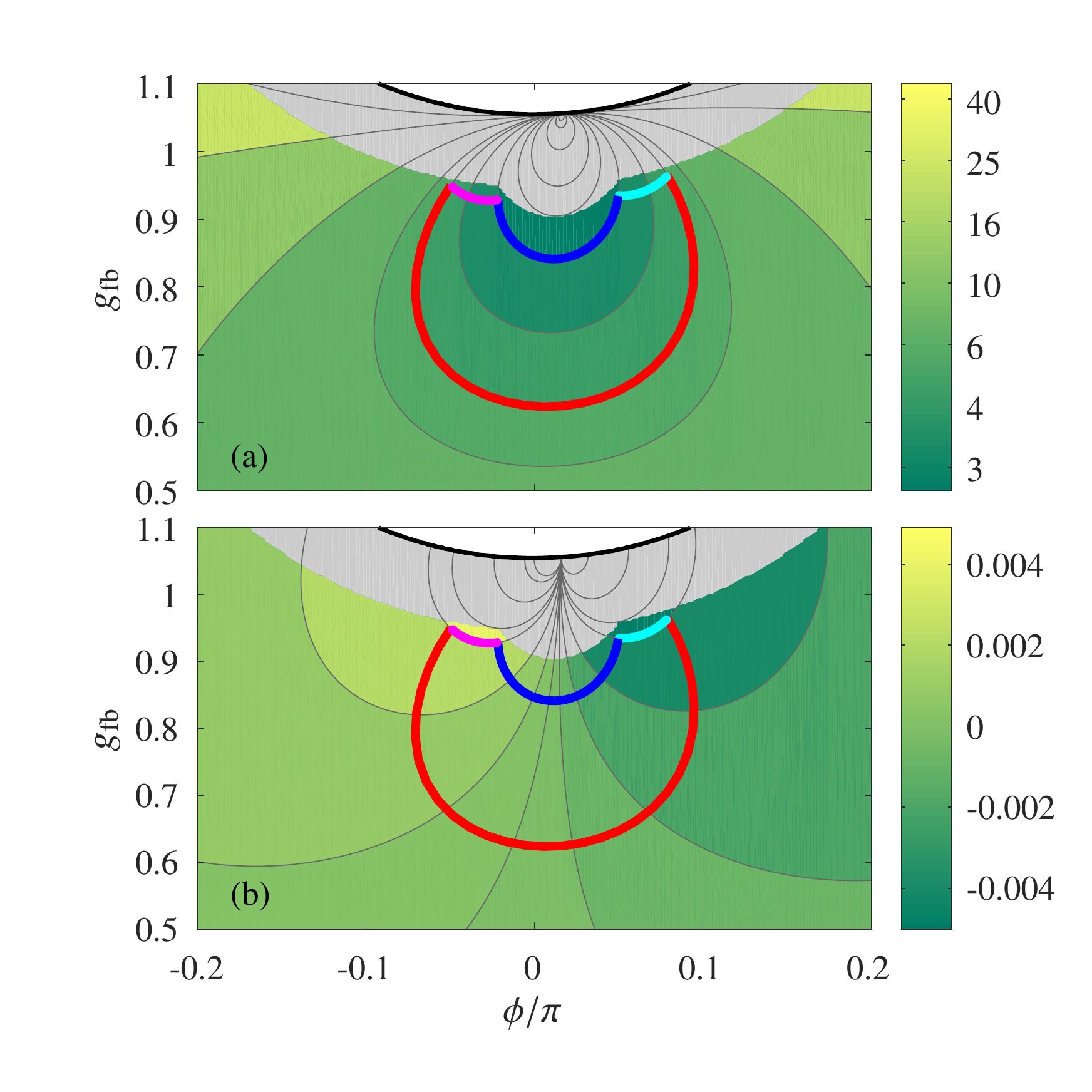}\\
\caption{As in Fig.~\ref{Fig4} with $\kappa=0.1~\omega_\mm$, $G=0.01~\omega_\mm$, $T_{\rm hot}=6$~K, $T_{\rm cold}=3$~K, $\Delta_\mm\al{h}=0.004~\omega_\mm$ and $\Delta_\mm\al{l}=-0.004~\omega_\mm$.
In this case the resulting efficiency (with $t_{\rm tot}=40000~\omega_\mm^{-1}$ and $r_T=0.5$) is $\eta= 0.01$.
}
\label{Fig11}
\end{figure}
\begin{figure}[!t]
\centering
\includegraphics[width=\columnwidth]{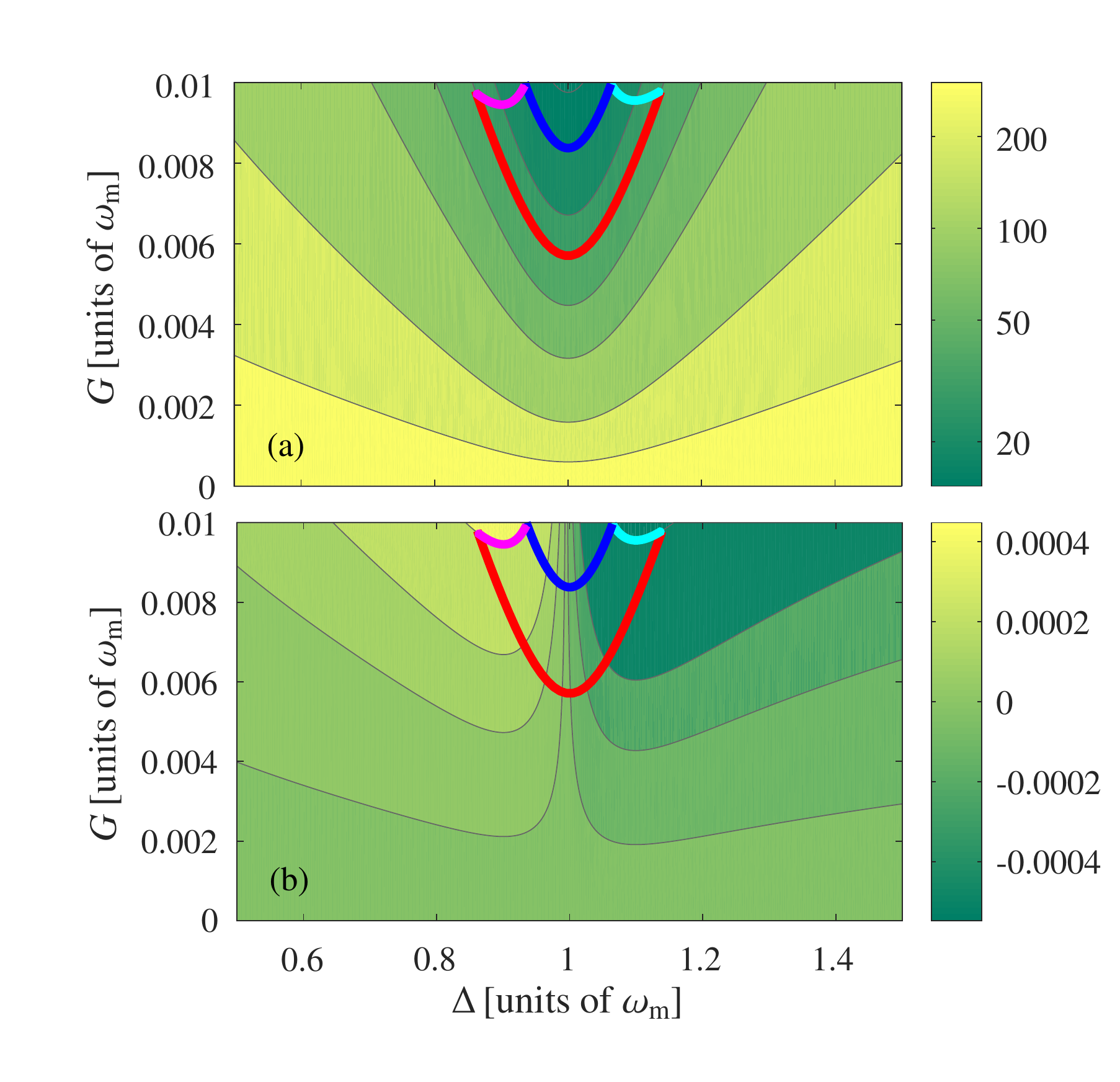}\\
\caption{As in Fig.~\ref{Fig7} with $\kappa=0.1~\omega_\mm$,
$T_{\rm hot}=40$~K, $T_{\rm cold}=20$~K, $\Delta_\mm\al{h}=0.0004~\omega_\mm$ and $\Delta_\mm\al{l}=-0.0005~\omega_\mm$.
In this case the resulting efficiency (with $t_{\rm tot}=60000\omega_\mm^{-1}$ and $r_T=0.5$) is $\eta= 0.001$.
}
\label{Fig12}
\end{figure}
\begin{figure}[t]
\centering
\includegraphics[width=\columnwidth]{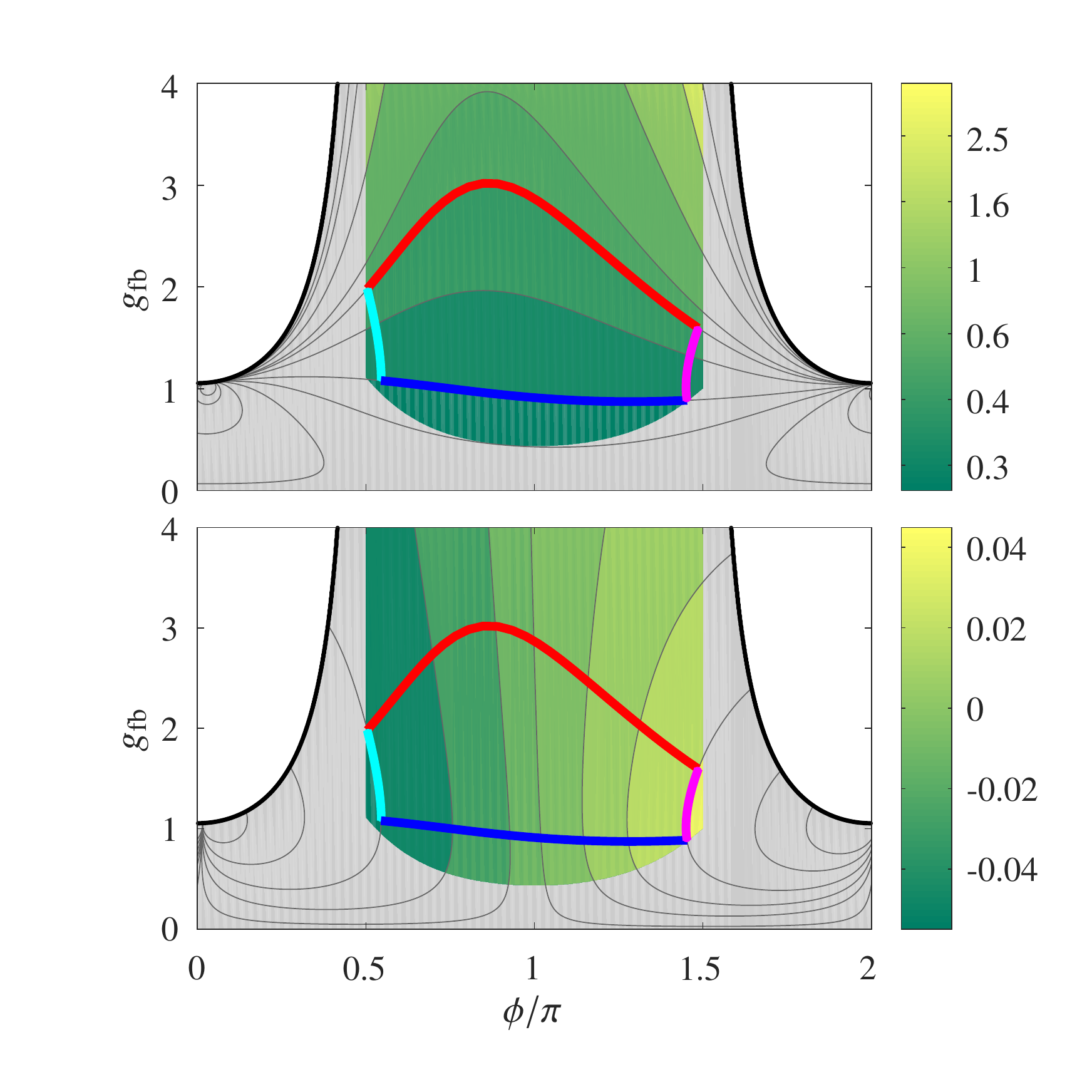}\\
\caption{As in Fig.~\ref{Fig4} with $\kappa=0.1~\omega_\mm$, $G=0.1~\omega_\mm$, $T_{\rm hot}=0.7$~K, $T_{\rm cold}=0.3$~K, $\Delta_\mm\al{h}=0.035~\omega_\mm$ and $\Delta_\mm\al{l}=-0.04~\omega_\mm$.
In this case the resulting efficiency (with $t_{\rm tot}=2000~\omega_\mm^{-1}$ and $r_T=0.5$) is $\eta= 0.06$.
}
\label{Fig13}
\end{figure}
\begin{figure}[t]
\centering
\includegraphics[width=\columnwidth]{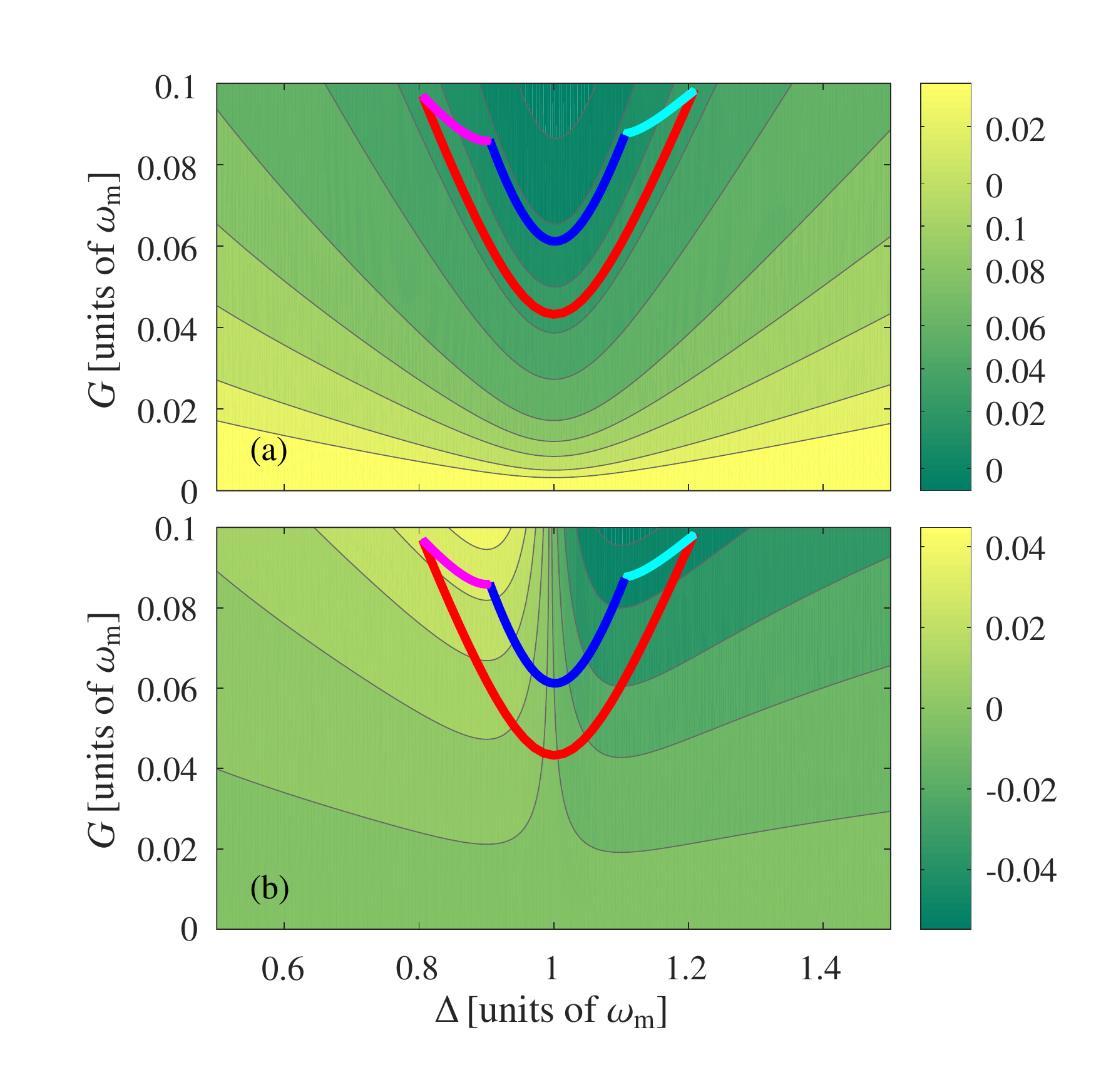}\\
\caption{As in Fig.~\ref{Fig7} with $\kappa=0.1~\omega_\mm$, $T_{\rm hot}=1$~K, $T_{\rm cold}=0.35$~K, $\Delta_\mm\al{h}=0.03~\omega_\mm$, $\Delta_\mm\al{l}=-0.035~\omega_\mm$.
In this case the resulting efficiency (with $t_{\rm tot}=2000~\omega_\mm^{-1}$ and $r_T=0.5$) is $\eta= 0.06$.
}
\label{Fig14}
\end{figure}

\subsection{The power spectrum of the cavity field}
\label{power-spectrum}
The expression for the power spectrum of the cavity quadrature $S_{X_0}(\omega)$ reported in Eq.~\rp{SX0} can be computed as follows. 
The operators of the cavity field, obtained solving Eq.~\rp{dotaeff}, are
\begin{eqnarray}
\tilde a(\omega)\Bigr|_{G=0}&=&\frac{\sqrt{2\,\kappa_\eff}\ \tilde\chi(\omega)}{1-\abs{\mu}^2\,\tilde\chi(\omega)\,\tilde\chi(-\omega)^*}
\\&&\times
\pq{ a_{\inn,\eff}(\omega)+\mu^*\,\tilde\chi(-\omega)^*\ a_{\inn,\eff}\da(\omega) }
\nn\\
\tilde a\da(\omega)\Bigr|_{G=0}&=&\frac{\sqrt{2\,\kappa_\eff}\ \tilde\chi(-\omega)^*}{1-\abs{\mu}^2\,\tilde\chi(\omega)\,\tilde\chi(-\omega)^*}
\nn\\&&\times
\pq{ a_{\inn,\eff}\da(\omega)+\mu\,\tilde\chi(\omega)\ a_{\inn,\eff}(\omega) }\ .
\nn
\end{eqnarray}
Hence, the corresponding expression for the quadrature operator is
\begin{eqnarray}\label{tildeX0App}
\tilde X_0(\omega)=\sqrt{2\,\kappa_\eff}\pq{\Lambda(\omega)\ a_{\inn,\eff}(\omega)+\Lambda(-\omega)^*\ a_{\inn,\eff}\da(\omega) }\ ,
\nn\\
\end{eqnarray}
where the cavity response function $\Lambda(\omega)$ is defined in Eq.~\rp{Lambda}.
Finally, Eq.~\rp{SX0} is obtained combining Eqs.~\rp{XXS}, \rp{tildeX0App} and  the correlations of the input noise in Fourier space  
\begin{eqnarray}
\av{a_{\inn,\eff}\da(\omega)\ a_{\inn,\eff}(\omega')} &=& n_\eff\ \delta(\omega+\omega') , \\ 
\av{a_{\inn,\eff}(\omega)\ a_{\inn,\eff}\da(\omega')} &=& (n_\eff+1)\ \delta(\omega+\omega') , \\ 
\av{a_{\inn,\eff}(\omega)\ a_{\inn,\eff}(\omega')} &=&m_\eff\ \delta(\omega+\omega') , \\ 
\av{a_{\inn,\eff}\da(\omega)\ a_{\inn,\eff}\da(\omega')} &=&m_\eff^*\ \delta(\omega+\omega') . 
\end{eqnarray}

\section{The resolved sideband limit}
\label{App-rsl}

In the main text we have studied an optomechanical system in the unresolved sideband regime ($\kappa=2\,\omega_\mm$) where our approach gives the largest efficiency.
% in this regime.
%
In this appendix, for completeness, we derive a few results in the resolved sideband limit, where $\kappa_\eff\ll\omega_\mm$. 
In this case the engine efficiency is reduced. 
The results that we report hereafter are obtained for a ratio $T_{\rm hot}/T_{\rm cold}=2$. Instead, the values of $\Delta_\mm\al{h}$ and $\Delta_\mm\al{l}$ are chosen in order to optimize
the corresponding engine efficiencies.

Specifically, here we consider $\kappa=0.1\omega_\mm$.
The results in Figs.~\ref{Fig11} and \ref{Fig12} are obtained with parameters consistent with those of the experiments 
reported in Refs.~\cite{rossi2017,Kralj2017,rossi2018}. 
Figure~\ref{Fig11} is obtained including the feedback, whereas Fig.~\ref{Fig12} is without feedback. We note that the corresponding engine efficiencies are very low, however also in this case we find a strong enhancement due to the feedback.

In Figs.~\ref{Fig13} and \ref{Fig14} instead we have used larger optomechanical couplings (up to $G=\kappa=0.1~\omega_\mm$). In this case the system is at the limit of validity of the adiabatic elimination, and the feedback can not be fully  exploited to enhance the performance of the engine. In fact, the possibility of extending the variability of the optomechanical parameters is permitted by the ability to reduce also the effective cavity linewidth (see Eq.~\rp{kappaeff-Deltaeff})~\cite{zippilli2018}. In this case, however, the cavity linewidth cannot be further reduced and the efficiencies  corresponding to the thermodynamic cycles with (Fig.~\ref{Fig13}) and without (Fig.~\ref{Fig14})  feedback are comparable. 

%%%%%%%%%%%%%%%%%%%%%%%%%%%%%%%%%%%%%%%%%%%%%%%%%%%%%%%%%%%%%%%%%%%%%%


%\bibliography{Quantum-heat-engine.bib}

\begin{thebibliography}{32}%
\makeatletter
\providecommand \@ifxundefined [1]{%
 \@ifx{#1\undefined}
}%
\providecommand \@ifnum [1]{%
 \ifnum #1\expandafter \@firstoftwo
 \else \expandafter \@secondoftwo
 \fi
}%
\providecommand \@ifx [1]{%
 \ifx #1\expandafter \@firstoftwo
 \else \expandafter \@secondoftwo
 \fi
}%
\providecommand \natexlab [1]{#1}%
\providecommand \enquote  [1]{``#1''}%
\providecommand \bibnamefont  [1]{#1}%
\providecommand \bibfnamefont [1]{#1}%
\providecommand \citenamefont [1]{#1}%
\providecommand \href@noop [0]{\@secondoftwo}%
\providecommand \href [0]{\begingroup \@sanitize@url \@href}%
\providecommand \@href[1]{\@@startlink{#1}\@@href}%
\providecommand \@@href[1]{\endgroup#1\@@endlink}%
\providecommand \@sanitize@url [0]{\catcode `\\12\catcode `\$12\catcode
  `\&12\catcode `\#12\catcode `\^12\catcode `\_12\catcode `\%12\relax}%
\providecommand \@@startlink[1]{}%
\providecommand \@@endlink[0]{}%
\providecommand \url  [0]{\begingroup\@sanitize@url \@url }%
\providecommand \@url [1]{\endgroup\@href {#1}{\urlprefix }}%
\providecommand \urlprefix  [0]{URL }%
\providecommand \Eprint [0]{\href }%
\providecommand \doibase [0]{http://dx.doi.org/}%
\providecommand \selectlanguage [0]{\@gobble}%
\providecommand \bibinfo  [0]{\@secondoftwo}%
\providecommand \bibfield  [0]{\@secondoftwo}%
\providecommand \translation [1]{[#1]}%
\providecommand \BibitemOpen [0]{}%
\providecommand \bibitemStop [0]{}%
\providecommand \bibitemNoStop [0]{.\EOS\space}%
\providecommand \EOS [0]{\spacefactor3000\relax}%
\providecommand \BibitemShut  [1]{\csname bibitem#1\endcsname}%
\let\auto@bib@innerbib\@empty
%</preamble>
\bibitem [{\citenamefont {Benenti}\ \emph {et~al.}(2017)\citenamefont
  {Benenti}, \citenamefont {Casati}, \citenamefont {Saito},\ and\ \citenamefont
  {Whitney}}]{benenti2017}%
  \BibitemOpen
  \bibfield  {author} {\bibinfo {author} {\bibfnamefont {Giuliano}\
  \bibnamefont {Benenti}}, \bibinfo {author} {\bibfnamefont {Giulio}\
  \bibnamefont {Casati}}, \bibinfo {author} {\bibfnamefont {Keiji}\
  \bibnamefont {Saito}}, \ and\ \bibinfo {author} {\bibfnamefont {Robert~S.}\
  \bibnamefont {Whitney}},\ }\bibfield  {title} {\enquote {\bibinfo {title}
  {Fundamental aspects of steady-state conversion of heat to work at the
  nanoscale},}\ }\href {\doibase 10.1016/j.physrep.2017.05.008} {\bibfield
  {journal} {\bibinfo  {journal} {Physics Reports}\ }\textbf {\bibinfo {volume}
  {694}},\ \bibinfo {pages} {1--124} (\bibinfo {year} {2017})}\BibitemShut
  {NoStop}%
\bibitem [{\citenamefont {Binder}\ \emph {et~al.}(2018)\citenamefont {Binder},
  \citenamefont {Correa}, \citenamefont {Gogolin}, \citenamefont {Anders},\
  and\ \citenamefont {Adesso}}]{binder2018}%
  \BibitemOpen
  \bibinfo {editor} {\bibfnamefont {Felix}\ \bibnamefont {Binder}}, \bibinfo
  {editor} {\bibfnamefont {Luis~A.}\ \bibnamefont {Correa}}, \bibinfo {editor}
  {\bibfnamefont {Christian}\ \bibnamefont {Gogolin}}, \bibinfo {editor}
  {\bibfnamefont {Janet}\ \bibnamefont {Anders}}, \ and\ \bibinfo {editor}
  {\bibfnamefont {Gerardo}\ \bibnamefont {Adesso}},\ eds.,\ \href {\doibase
  10.1007/978-3-319-99046-0} {\emph {\bibinfo {title} {Thermodynamics in the
  {{Quantum Regime}}: {{Fundamental Aspects}} and {{New Directions}}}}},\
  Fundamental {{Theories}} of {{Physics}}\ (\bibinfo  {publisher} {{Springer
  International Publishing}},\ \bibinfo {address} {{Cham}},\ \bibinfo {year}
  {2018})\BibitemShut {NoStop}%
\bibitem [{\citenamefont {Aspelmeyer}\ \emph {et~al.}(2014)\citenamefont
  {Aspelmeyer}, \citenamefont {Kippenberg},\ and\ \citenamefont
  {Marquardt}}]{aspelmeyer2014}%
  \BibitemOpen
  \bibfield  {author} {\bibinfo {author} {\bibfnamefont {Markus}\ \bibnamefont
  {Aspelmeyer}}, \bibinfo {author} {\bibfnamefont {Tobias~J.}\ \bibnamefont
  {Kippenberg}}, \ and\ \bibinfo {author} {\bibfnamefont {Florian}\
  \bibnamefont {Marquardt}},\ }\bibfield  {title} {\enquote {\bibinfo {title}
  {Cavity optomechanics},}\ }\href {\doibase 10.1103/RevModPhys.86.1391}
  {\bibfield  {journal} {\bibinfo  {journal} {Rev. Mod. Phys.}\ }\textbf
  {\bibinfo {volume} {86}},\ \bibinfo {pages} {1391--1452} (\bibinfo {year}
  {2014})}\BibitemShut {NoStop}%
\bibitem [{\citenamefont {Bowen}\ and\ \citenamefont
  {Milburn}(2015)}]{bowen2015}%
  \BibitemOpen
  \bibfield  {author} {\bibinfo {author} {\bibfnamefont {Warwick~P.}\
  \bibnamefont {Bowen}}\ and\ \bibinfo {author} {\bibfnamefont {Gerard~J.}\
  \bibnamefont {Milburn}},\ }\href@noop {} {\emph {\bibinfo {title} {Quantum
  {{Optomechanics}}}}}\ (\bibinfo  {publisher} {{Taylor \& Francis}},\ \bibinfo
  {year} {2015})\BibitemShut {NoStop}%
\bibitem [{\citenamefont {Zhang}\ \emph
  {et~al.}(2014{\natexlab{a}})\citenamefont {Zhang}, \citenamefont {Bariani},\
  and\ \citenamefont {Meystre}}]{zhang2014a}%
  \BibitemOpen
  \bibfield  {author} {\bibinfo {author} {\bibfnamefont {Keye}\ \bibnamefont
  {Zhang}}, \bibinfo {author} {\bibfnamefont {Francesco}\ \bibnamefont
  {Bariani}}, \ and\ \bibinfo {author} {\bibfnamefont {Pierre}\ \bibnamefont
  {Meystre}},\ }\bibfield  {title} {\enquote {\bibinfo {title} {Quantum
  {{Optomechanical Heat Engine}}},}\ }\href {\doibase
  10.1103/PhysRevLett.112.150602} {\bibfield  {journal} {\bibinfo  {journal}
  {Phys. Rev. Lett.}\ }\textbf {\bibinfo {volume} {112}},\ \bibinfo {pages}
  {150602} (\bibinfo {year} {2014}{\natexlab{a}})}\BibitemShut {NoStop}%
\bibitem [{\citenamefont {Zhang}\ \emph
  {et~al.}(2014{\natexlab{b}})\citenamefont {Zhang}, \citenamefont {Bariani},\
  and\ \citenamefont {Meystre}}]{zhang2014}%
  \BibitemOpen
  \bibfield  {author} {\bibinfo {author} {\bibfnamefont {Keye}\ \bibnamefont
  {Zhang}}, \bibinfo {author} {\bibfnamefont {Francesco}\ \bibnamefont
  {Bariani}}, \ and\ \bibinfo {author} {\bibfnamefont {Pierre}\ \bibnamefont
  {Meystre}},\ }\bibfield  {title} {\enquote {\bibinfo {title} {Theory of an
  optomechanical quantum heat engine},}\ }\href {\doibase
  10.1103/PhysRevA.90.023819} {\bibfield  {journal} {\bibinfo  {journal} {Phys.
  Rev. A}\ }\textbf {\bibinfo {volume} {90}},\ \bibinfo {pages} {023819}
  (\bibinfo {year} {2014}{\natexlab{b}})}\BibitemShut {NoStop}%
\bibitem [{\citenamefont {Dong}\ \emph
  {et~al.}(2015{\natexlab{a}})\citenamefont {Dong}, \citenamefont {Zhang},
  \citenamefont {Bariani},\ and\ \citenamefont {Meystre}}]{dong2015}%
  \BibitemOpen
  \bibfield  {author} {\bibinfo {author} {\bibfnamefont {Ying}\ \bibnamefont
  {Dong}}, \bibinfo {author} {\bibfnamefont {Keye}\ \bibnamefont {Zhang}},
  \bibinfo {author} {\bibfnamefont {Francesco}\ \bibnamefont {Bariani}}, \ and\
  \bibinfo {author} {\bibfnamefont {Pierre}\ \bibnamefont {Meystre}},\
  }\bibfield  {title} {\enquote {\bibinfo {title} {Work measurement in an
  optomechanical quantum heat engine},}\ }\href {\doibase
  10.1103/PhysRevA.92.033854} {\bibfield  {journal} {\bibinfo  {journal} {Phys.
  Rev. A}\ }\textbf {\bibinfo {volume} {92}},\ \bibinfo {pages} {033854}
  (\bibinfo {year} {2015}{\natexlab{a}})}\BibitemShut {NoStop}%
\bibitem [{\citenamefont {Dong}\ \emph
  {et~al.}(2015{\natexlab{b}})\citenamefont {Dong}, \citenamefont {Bariani},\
  and\ \citenamefont {Meystre}}]{dong2015a}%
  \BibitemOpen
  \bibfield  {author} {\bibinfo {author} {\bibfnamefont {Ying}\ \bibnamefont
  {Dong}}, \bibinfo {author} {\bibfnamefont {F.}~\bibnamefont {Bariani}}, \
  and\ \bibinfo {author} {\bibfnamefont {P.}~\bibnamefont {Meystre}},\
  }\bibfield  {title} {\enquote {\bibinfo {title} {Phonon {{Cooling}} by an
  {{Optomechanical Heat Pump}}},}\ }\href {\doibase
  10.1103/PhysRevLett.115.223602} {\bibfield  {journal} {\bibinfo  {journal}
  {Phys. Rev. Lett.}\ }\textbf {\bibinfo {volume} {115}},\ \bibinfo {pages}
  {223602} (\bibinfo {year} {2015}{\natexlab{b}})}\BibitemShut {NoStop}%
\bibitem [{\citenamefont {Dechant}\ \emph {et~al.}(2015)\citenamefont
  {Dechant}, \citenamefont {Kiesel},\ and\ \citenamefont
  {Lutz}}]{dechant2015a}%
  \BibitemOpen
  \bibfield  {author} {\bibinfo {author} {\bibfnamefont {Andreas}\ \bibnamefont
  {Dechant}}, \bibinfo {author} {\bibfnamefont {Nikolai}\ \bibnamefont
  {Kiesel}}, \ and\ \bibinfo {author} {\bibfnamefont {Eric}\ \bibnamefont
  {Lutz}},\ }\bibfield  {title} {\enquote {\bibinfo {title} {All-{{Optical
  Nanomechanical Heat Engine}}},}\ }\href {\doibase
  10.1103/PhysRevLett.114.183602} {\bibfield  {journal} {\bibinfo  {journal}
  {Phys. Rev. Lett.}\ }\textbf {\bibinfo {volume} {114}},\ \bibinfo {pages}
  {183602} (\bibinfo {year} {2015})}\BibitemShut {NoStop}%
\bibitem [{\citenamefont {Mari}\ \emph {et~al.}(2015)\citenamefont {Mari},
  \citenamefont {Farace},\ and\ \citenamefont {Giovannetti}}]{mari2015}%
  \BibitemOpen
  \bibfield  {author} {\bibinfo {author} {\bibfnamefont {A.}~\bibnamefont
  {Mari}}, \bibinfo {author} {\bibfnamefont {A.}~\bibnamefont {Farace}}, \ and\
  \bibinfo {author} {\bibfnamefont {V.}~\bibnamefont {Giovannetti}},\
  }\bibfield  {title} {\enquote {\bibinfo {title} {Quantum optomechanical
  piston engines powered by heat},}\ }\href {\doibase
  10.1088/0953-4075/48/17/175501} {\bibfield  {journal} {\bibinfo  {journal}
  {J. Phys. B: At. Mol. Opt. Phys.}\ }\textbf {\bibinfo {volume} {48}},\
  \bibinfo {pages} {175501} (\bibinfo {year} {2015})}\BibitemShut {NoStop}%
\bibitem [{\citenamefont {{Gelbwaser-Klimovsky}}\ and\ \citenamefont
  {Kurizki}(2015)}]{gelbwaser-klimovsky2015a}%
  \BibitemOpen
  \bibfield  {author} {\bibinfo {author} {\bibfnamefont {D.}~\bibnamefont
  {{Gelbwaser-Klimovsky}}}\ and\ \bibinfo {author} {\bibfnamefont
  {G.}~\bibnamefont {Kurizki}},\ }\bibfield  {title} {\enquote {\bibinfo
  {title} {Work extraction from heat-powered quantized optomechanical
  setups},}\ }\href {\doibase 10.1038/srep07809} {\bibfield  {journal}
  {\bibinfo  {journal} {Sci. Rep.}\ }\textbf {\bibinfo {volume} {5}},\ \bibinfo
  {pages} {07809} (\bibinfo {year} {2015})}\BibitemShut {NoStop}%
\bibitem [{\citenamefont {Bathaee}\ and\ \citenamefont
  {Bahrampour}(2016)}]{Bathaee2016}%
  \BibitemOpen
  \bibfield  {author} {\bibinfo {author} {\bibfnamefont {M.}~\bibnamefont
  {Bathaee}}\ and\ \bibinfo {author} {\bibfnamefont {A.~R.}\ \bibnamefont
  {Bahrampour}},\ }\bibfield  {title} {\enquote {\bibinfo {title} {Optimal
  control of the power adiabatic stroke of an optomechanical heat engine},}\
  }\href {\doibase 10.1103/PhysRevE.94.022141} {\bibfield  {journal} {\bibinfo
  {journal} {Phys. Rev. E}\ }\textbf {\bibinfo {volume} {94}},\ \bibinfo
  {pages} {022141} (\bibinfo {year} {2016})}\BibitemShut {NoStop}%
\bibitem [{\citenamefont {Zhang}\ and\ \citenamefont
  {Zhang}(2017)}]{zhang2017}%
  \BibitemOpen
  \bibfield  {author} {\bibinfo {author} {\bibfnamefont {Keye}\ \bibnamefont
  {Zhang}}\ and\ \bibinfo {author} {\bibfnamefont {Weiping}\ \bibnamefont
  {Zhang}},\ }\bibfield  {title} {\enquote {\bibinfo {title} {Quantum
  optomechanical straight-twin engine},}\ }\href {\doibase
  10.1103/PhysRevA.95.053870} {\bibfield  {journal} {\bibinfo  {journal} {Phys.
  Rev. A}\ }\textbf {\bibinfo {volume} {95}},\ \bibinfo {pages} {053870}
  (\bibinfo {year} {2017})}\BibitemShut {NoStop}%
\bibitem [{\citenamefont {Bennett}\ \emph {et~al.}(2020)\citenamefont
  {Bennett}, \citenamefont {Madsen}, \citenamefont {{Rubinsztein-Dunlop}},\
  and\ \citenamefont {Bowen}}]{bennett2020a}%
  \BibitemOpen
  \bibfield  {author} {\bibinfo {author} {\bibfnamefont {James~S.}\
  \bibnamefont {Bennett}}, \bibinfo {author} {\bibfnamefont {Lars~S.}\
  \bibnamefont {Madsen}}, \bibinfo {author} {\bibfnamefont {Halina}\
  \bibnamefont {{Rubinsztein-Dunlop}}}, \ and\ \bibinfo {author} {\bibfnamefont
  {Warwick~P.}\ \bibnamefont {Bowen}},\ }\bibfield  {title} {\enquote {\bibinfo
  {title} {A quantum heat machine from fast optomechanics},}\ }\href {\doibase
  10.1088/1367-2630/abb73f} {\bibfield  {journal} {\bibinfo  {journal} {New J.
  Phys.}\ }\textbf {\bibinfo {volume} {22}},\ \bibinfo {pages} {103028}
  (\bibinfo {year} {2020})}\BibitemShut {NoStop}%
\bibitem [{\citenamefont {Naseem}\ and\ \citenamefont {M{\"u}stecaplio{\u
  g}lu}(2019)}]{naseem2019}%
  \BibitemOpen
  \bibfield  {author} {\bibinfo {author} {\bibfnamefont {M.~Tahir}\
  \bibnamefont {Naseem}}\ and\ \bibinfo {author} {\bibfnamefont
  {{\"O}zg{\"u}r~E.}\ \bibnamefont {M{\"u}stecaplio{\u g}lu}},\ }\bibfield
  {title} {\enquote {\bibinfo {title} {Quantum heat engine with a quadratically
  coupled optomechanical system},}\ }\href {\doibase 10.1364/JOSAB.36.003000}
  {\bibfield  {journal} {\bibinfo  {journal} {J. Opt. Soc. Am. B}\ }\textbf
  {\bibinfo {volume} {36}},\ \bibinfo {pages} {3000--3008} (\bibinfo {year}
  {2019})}\BibitemShut {NoStop}%
\bibitem [{\citenamefont {Abari}\ \emph {et~al.}(2019)\citenamefont {Abari},
  \citenamefont {Angelis}, \citenamefont {Zippilli},\ and\ \citenamefont
  {Vitali}}]{abari2019}%
  \BibitemOpen
  \bibfield  {author} {\bibinfo {author} {\bibfnamefont {Najmeh~Etehadi}\
  \bibnamefont {Abari}}, \bibinfo {author} {\bibfnamefont {Giulia Vittoria~De}\
  \bibnamefont {Angelis}}, \bibinfo {author} {\bibfnamefont {Stefano}\
  \bibnamefont {Zippilli}}, \ and\ \bibinfo {author} {\bibfnamefont {David}\
  \bibnamefont {Vitali}},\ }\bibfield  {title} {\enquote {\bibinfo {title} {An
  optomechanical heat engine with feedback-controlled in-loop light},}\ }\href
  {\doibase 10.1088/1367-2630/ab41e7} {\bibfield  {journal} {\bibinfo
  {journal} {New J. Phys.}\ }\textbf {\bibinfo {volume} {21}},\ \bibinfo
  {pages} {093051} (\bibinfo {year} {2019})}\BibitemShut {NoStop}%
\bibitem [{\citenamefont {Zippilli}\ \emph {et~al.}(2018)\citenamefont
  {Zippilli}, \citenamefont {Kralj}, \citenamefont {Rossi}, \citenamefont
  {Di~Giuseppe},\ and\ \citenamefont {Vitali}}]{zippilli2018}%
  \BibitemOpen
  \bibfield  {author} {\bibinfo {author} {\bibfnamefont {Stefano}\ \bibnamefont
  {Zippilli}}, \bibinfo {author} {\bibfnamefont {Nenad}\ \bibnamefont {Kralj}},
  \bibinfo {author} {\bibfnamefont {Massimiliano}\ \bibnamefont {Rossi}},
  \bibinfo {author} {\bibfnamefont {Giovanni}\ \bibnamefont {Di~Giuseppe}}, \
  and\ \bibinfo {author} {\bibfnamefont {David}\ \bibnamefont {Vitali}},\
  }\bibfield  {title} {\enquote {\bibinfo {title} {Cavity optomechanics with
  feedback-controlled in-loop light},}\ }\href {\doibase
  10.1103/PhysRevA.98.023828} {\bibfield  {journal} {\bibinfo  {journal} {Phys.
  Rev. A}\ }\textbf {\bibinfo {volume} {98}},\ \bibinfo {pages} {023828}
  (\bibinfo {year} {2018})}\BibitemShut {NoStop}%
\bibitem [{\citenamefont {Rossi}\ \emph {et~al.}(2017)\citenamefont {Rossi},
  \citenamefont {Kralj}, \citenamefont {Zippilli}, \citenamefont {Natali},
  \citenamefont {Borrielli}, \citenamefont {Pandraud}, \citenamefont {Serra},
  \citenamefont {Di~Giuseppe},\ and\ \citenamefont {Vitali}}]{rossi2017}%
  \BibitemOpen
  \bibfield  {author} {\bibinfo {author} {\bibfnamefont {Massimiliano}\
  \bibnamefont {Rossi}}, \bibinfo {author} {\bibfnamefont {Nenad}\ \bibnamefont
  {Kralj}}, \bibinfo {author} {\bibfnamefont {Stefano}\ \bibnamefont
  {Zippilli}}, \bibinfo {author} {\bibfnamefont {Riccardo}\ \bibnamefont
  {Natali}}, \bibinfo {author} {\bibfnamefont {Antonio}\ \bibnamefont
  {Borrielli}}, \bibinfo {author} {\bibfnamefont {Gregory}\ \bibnamefont
  {Pandraud}}, \bibinfo {author} {\bibfnamefont {Enrico}\ \bibnamefont
  {Serra}}, \bibinfo {author} {\bibfnamefont {Giovanni}\ \bibnamefont
  {Di~Giuseppe}}, \ and\ \bibinfo {author} {\bibfnamefont {David}\ \bibnamefont
  {Vitali}},\ }\bibfield  {title} {\enquote {\bibinfo {title} {Enhancing
  {{Sideband Cooling}} by {{Feedback}}-{{Controlled Light}}},}\ }\href
  {\doibase 10.1103/PhysRevLett.119.123603} {\bibfield  {journal} {\bibinfo
  {journal} {Phys. Rev. Lett.}\ }\textbf {\bibinfo {volume} {119}},\ \bibinfo
  {pages} {123603} (\bibinfo {year} {2017})}\BibitemShut {NoStop}%
\bibitem [{\citenamefont {Kralj}\ \emph {et~al.}(2017)\citenamefont {Kralj},
  \citenamefont {Rossi}, \citenamefont {Zippilli}, \citenamefont {Natali},
  \citenamefont {Borrielli}, \citenamefont {Pandraud}, \citenamefont {Serra},
  \citenamefont {Giuseppe},\ and\ \citenamefont {Vitali}}]{Kralj2017}%
  \BibitemOpen
  \bibfield  {author} {\bibinfo {author} {\bibfnamefont {Nenad}\ \bibnamefont
  {Kralj}}, \bibinfo {author} {\bibfnamefont {Massimiliano}\ \bibnamefont
  {Rossi}}, \bibinfo {author} {\bibfnamefont {Stefano}\ \bibnamefont
  {Zippilli}}, \bibinfo {author} {\bibfnamefont {Riccardo}\ \bibnamefont
  {Natali}}, \bibinfo {author} {\bibfnamefont {Antonio}\ \bibnamefont
  {Borrielli}}, \bibinfo {author} {\bibfnamefont {Gregory}\ \bibnamefont
  {Pandraud}}, \bibinfo {author} {\bibfnamefont {Enrico}\ \bibnamefont
  {Serra}}, \bibinfo {author} {\bibfnamefont {Giovanni~Di}\ \bibnamefont
  {Giuseppe}}, \ and\ \bibinfo {author} {\bibfnamefont {David}\ \bibnamefont
  {Vitali}},\ }\bibfield  {title} {\enquote {\bibinfo {title} {Enhancement of
  three-mode optomechanical interaction by feedback-controlled light},}\ }\href
  {\doibase 10.1088/2058-9565/aa7d7e} {\bibfield  {journal} {\bibinfo
  {journal} {Quantum Sci. Technol.}\ }\textbf {\bibinfo {volume} {2}},\
  \bibinfo {pages} {034014} (\bibinfo {year} {2017})}\BibitemShut {NoStop}%
\bibitem [{\citenamefont {Rossi}\ \emph {et~al.}(2018)\citenamefont {Rossi},
  \citenamefont {Kralj}, \citenamefont {Zippilli}, \citenamefont {Natali},
  \citenamefont {Borrielli}, \citenamefont {Pandraud}, \citenamefont {Serra},
  \citenamefont {Di~Giuseppe},\ and\ \citenamefont {Vitali}}]{rossi2018}%
  \BibitemOpen
  \bibfield  {author} {\bibinfo {author} {\bibfnamefont {Massimiliano}\
  \bibnamefont {Rossi}}, \bibinfo {author} {\bibfnamefont {Nenad}\ \bibnamefont
  {Kralj}}, \bibinfo {author} {\bibfnamefont {Stefano}\ \bibnamefont
  {Zippilli}}, \bibinfo {author} {\bibfnamefont {Riccardo}\ \bibnamefont
  {Natali}}, \bibinfo {author} {\bibfnamefont {Antonio}\ \bibnamefont
  {Borrielli}}, \bibinfo {author} {\bibfnamefont {Gregory}\ \bibnamefont
  {Pandraud}}, \bibinfo {author} {\bibfnamefont {Enrico}\ \bibnamefont
  {Serra}}, \bibinfo {author} {\bibfnamefont {Giovanni}\ \bibnamefont
  {Di~Giuseppe}}, \ and\ \bibinfo {author} {\bibfnamefont {David}\ \bibnamefont
  {Vitali}},\ }\bibfield  {title} {\enquote {\bibinfo {title} {Normal-{{Mode
  Splitting}} in a {{Weakly Coupled Optomechanical System}}},}\ }\href
  {\doibase 10.1103/PhysRevLett.120.073601} {\bibfield  {journal} {\bibinfo
  {journal} {Phys. Rev. Lett.}\ }\textbf {\bibinfo {volume} {120}},\ \bibinfo
  {pages} {073601} (\bibinfo {year} {2018})}\BibitemShut {NoStop}%
\bibitem [{\citenamefont {{Gelbwaser-Klimovsky}}\ and\ \citenamefont
  {{Aspuru-Guzik}}(2015)}]{gelbwaser-klimovsky2015b}%
  \BibitemOpen
  \bibfield  {author} {\bibinfo {author} {\bibfnamefont {David}\ \bibnamefont
  {{Gelbwaser-Klimovsky}}}\ and\ \bibinfo {author} {\bibfnamefont {Al{\'a}n}\
  \bibnamefont {{Aspuru-Guzik}}},\ }\bibfield  {title} {\enquote {\bibinfo
  {title} {Strongly {{Coupled Quantum Heat Machines}}},}\ }\href {\doibase
  10.1021/acs.jpclett.5b01404} {\bibfield  {journal} {\bibinfo  {journal} {J.
  Phys. Chem. Lett.}\ }\textbf {\bibinfo {volume} {6}},\ \bibinfo {pages}
  {3477--3482} (\bibinfo {year} {2015})}\BibitemShut {NoStop}%
\bibitem [{\citenamefont {Uzdin}\ \emph {et~al.}(2016)\citenamefont {Uzdin},
  \citenamefont {Levy},\ and\ \citenamefont {Kosloff}}]{uzdin2016}%
  \BibitemOpen
  \bibfield  {author} {\bibinfo {author} {\bibfnamefont {Raam}\ \bibnamefont
  {Uzdin}}, \bibinfo {author} {\bibfnamefont {Amikam}\ \bibnamefont {Levy}}, \
  and\ \bibinfo {author} {\bibfnamefont {Ronnie}\ \bibnamefont {Kosloff}},\
  }\bibfield  {title} {\enquote {\bibinfo {title} {Quantum {{Heat Machines
  Equivalence}}, {{Work Extraction}} beyond {{Markovianity}}, and {{Strong
  Coupling}} via {{Heat Exchangers}}},}\ }\href {\doibase 10.3390/e18040124}
  {\bibfield  {journal} {\bibinfo  {journal} {Entropy}\ }\textbf {\bibinfo
  {volume} {18}},\ \bibinfo {pages} {124} (\bibinfo {year} {2016})}\BibitemShut
  {NoStop}%
\bibitem [{\citenamefont {Wiedmann}\ \emph {et~al.}(2020)\citenamefont
  {Wiedmann}, \citenamefont {Stockburger},\ and\ \citenamefont
  {Ankerhold}}]{wiedmann2020a}%
  \BibitemOpen
  \bibfield  {author} {\bibinfo {author} {\bibfnamefont {M.}~\bibnamefont
  {Wiedmann}}, \bibinfo {author} {\bibfnamefont {J.~T.}\ \bibnamefont
  {Stockburger}}, \ and\ \bibinfo {author} {\bibfnamefont {J.}~\bibnamefont
  {Ankerhold}},\ }\bibfield  {title} {\enquote {\bibinfo {title}
  {Non-{{Markovian}} dynamics of a quantum heat engine: Out-of-equilibrium
  operation and thermal coupling control},}\ }\href {\doibase
  10.1088/1367-2630/ab725a} {\bibfield  {journal} {\bibinfo  {journal} {New J.
  Phys.}\ }\textbf {\bibinfo {volume} {22}},\ \bibinfo {pages} {033007}
  (\bibinfo {year} {2020})}\BibitemShut {NoStop}%
\bibitem [{\citenamefont {Newman}\ \emph {et~al.}(2020)\citenamefont {Newman},
  \citenamefont {Mintert},\ and\ \citenamefont {Nazir}}]{newman2020}%
  \BibitemOpen
  \bibfield  {author} {\bibinfo {author} {\bibfnamefont {David}\ \bibnamefont
  {Newman}}, \bibinfo {author} {\bibfnamefont {Florian}\ \bibnamefont
  {Mintert}}, \ and\ \bibinfo {author} {\bibfnamefont {Ahsan}\ \bibnamefont
  {Nazir}},\ }\bibfield  {title} {\enquote {\bibinfo {title} {Quantum limit to
  nonequilibrium heat-engine performance imposed by strong system-reservoir
  coupling},}\ }\href {\doibase 10.1103/PhysRevE.101.052129} {\bibfield
  {journal} {\bibinfo  {journal} {Phys. Rev. E}\ }\textbf {\bibinfo {volume}
  {101}},\ \bibinfo {pages} {052129} (\bibinfo {year} {2020})}\BibitemShut
  {NoStop}%
\bibitem [{\citenamefont {Abah}\ and\ \citenamefont {Lutz}(2014)}]{abah2014a}%
  \BibitemOpen
  \bibfield  {author} {\bibinfo {author} {\bibfnamefont {Obinna}\ \bibnamefont
  {Abah}}\ and\ \bibinfo {author} {\bibfnamefont {Eric}\ \bibnamefont {Lutz}},\
  }\bibfield  {title} {\enquote {\bibinfo {title} {Efficiency of heat engines
  coupled to nonequilibrium reservoirs},}\ }\href {\doibase
  10.1209/0295-5075/106/20001} {\bibfield  {journal} {\bibinfo  {journal}
  {EPL}\ }\textbf {\bibinfo {volume} {106}},\ \bibinfo {pages} {20001}
  (\bibinfo {year} {2014})}\BibitemShut {NoStop}%
\bibitem [{\citenamefont {Niedenzu}\ \emph {et~al.}(2016)\citenamefont
  {Niedenzu}, \citenamefont {{Gelbwaser-Klimovsky}}, \citenamefont {Kofman},\
  and\ \citenamefont {Kurizki}}]{niedenzu2016}%
  \BibitemOpen
  \bibfield  {author} {\bibinfo {author} {\bibfnamefont {Wolfgang}\
  \bibnamefont {Niedenzu}}, \bibinfo {author} {\bibfnamefont {David}\
  \bibnamefont {{Gelbwaser-Klimovsky}}}, \bibinfo {author} {\bibfnamefont
  {Abraham~G.}\ \bibnamefont {Kofman}}, \ and\ \bibinfo {author} {\bibfnamefont
  {Gershon}\ \bibnamefont {Kurizki}},\ }\bibfield  {title} {\enquote {\bibinfo
  {title} {On the operation of machines powered by quantum non-thermal
  baths},}\ }\href {\doibase 10.1088/1367-2630/18/8/083012} {\bibfield
  {journal} {\bibinfo  {journal} {New J. Phys.}\ }\textbf {\bibinfo {volume}
  {18}},\ \bibinfo {pages} {083012} (\bibinfo {year} {2016})}\BibitemShut
  {NoStop}%
\bibitem [{\citenamefont {Campisi}\ \emph {et~al.}(2017)\citenamefont
  {Campisi}, \citenamefont {Pekola},\ and\ \citenamefont
  {Fazio}}]{campisi2017}%
  \BibitemOpen
  \bibfield  {author} {\bibinfo {author} {\bibfnamefont {Michele}\ \bibnamefont
  {Campisi}}, \bibinfo {author} {\bibfnamefont {Jukka}\ \bibnamefont {Pekola}},
  \ and\ \bibinfo {author} {\bibfnamefont {Rosario}\ \bibnamefont {Fazio}},\
  }\bibfield  {title} {\enquote {\bibinfo {title} {Feedback-controlled heat
  transport in quantum devices: Theory and solid-state experimental
  proposal},}\ }\href {\doibase 10.1088/1367-2630/aa6acb} {\bibfield  {journal}
  {\bibinfo  {journal} {New J. Phys.}\ }\textbf {\bibinfo {volume} {19}},\
  \bibinfo {pages} {053027} (\bibinfo {year} {2017})}\BibitemShut {NoStop}%
\bibitem [{\citenamefont {Potts}\ and\ \citenamefont
  {Samuelsson}(2018)}]{potts2018}%
  \BibitemOpen
  \bibfield  {author} {\bibinfo {author} {\bibfnamefont {Patrick~P.}\
  \bibnamefont {Potts}}\ and\ \bibinfo {author} {\bibfnamefont {Peter}\
  \bibnamefont {Samuelsson}},\ }\bibfield  {title} {\enquote {\bibinfo {title}
  {Detailed {{Fluctuation Relation}} for {{Arbitrary Measurement}} and
  {{Feedback Schemes}}},}\ }\href {\doibase 10.1103/PhysRevLett.121.210603}
  {\bibfield  {journal} {\bibinfo  {journal} {Phys. Rev. Lett.}\ }\textbf
  {\bibinfo {volume} {121}},\ \bibinfo {pages} {210603} (\bibinfo {year}
  {2018})}\BibitemShut {NoStop}%
\bibitem [{\citenamefont {Klaers}\ \emph {et~al.}(2017)\citenamefont {Klaers},
  \citenamefont {Faelt}, \citenamefont {Imamoglu},\ and\ \citenamefont
  {Togan}}]{klaers2017}%
  \BibitemOpen
  \bibfield  {author} {\bibinfo {author} {\bibfnamefont {Jan}\ \bibnamefont
  {Klaers}}, \bibinfo {author} {\bibfnamefont {Stefan}\ \bibnamefont {Faelt}},
  \bibinfo {author} {\bibfnamefont {Atac}\ \bibnamefont {Imamoglu}}, \ and\
  \bibinfo {author} {\bibfnamefont {Emre}\ \bibnamefont {Togan}},\ }\bibfield
  {title} {\enquote {\bibinfo {title} {Squeezed {{Thermal Reservoirs}} as a
  {{Resource}} for a {{Nanomechanical Engine}} beyond the {{Carnot Limit}}},}\
  }\href {\doibase 10.1103/PhysRevX.7.031044} {\bibfield  {journal} {\bibinfo
  {journal} {Phys. Rev. X}\ }\textbf {\bibinfo {volume} {7}},\ \bibinfo {pages}
  {031044} (\bibinfo {year} {2017})}\BibitemShut {NoStop}%
\bibitem [{\citenamefont {Abah}\ and\ \citenamefont {Lutz}(2017)}]{abah2017}%
  \BibitemOpen
  \bibfield  {author} {\bibinfo {author} {\bibfnamefont {Obinna}\ \bibnamefont
  {Abah}}\ and\ \bibinfo {author} {\bibfnamefont {Eric}\ \bibnamefont {Lutz}},\
  }\bibfield  {title} {\enquote {\bibinfo {title} {Energy efficient quantum
  machines},}\ }\href {\doibase 10.1209/0295-5075/118/40005} {\bibfield
  {journal} {\bibinfo  {journal} {EPL}\ }\textbf {\bibinfo {volume} {118}},\
  \bibinfo {pages} {40005} (\bibinfo {year} {2017})}\BibitemShut {NoStop}%
\bibitem [{\citenamefont {{Gomez-Marin}}\ \emph {et~al.}(2008)\citenamefont
  {{Gomez-Marin}}, \citenamefont {Schmiedl},\ and\ \citenamefont
  {Seifert}}]{gomez-marin2008}%
  \BibitemOpen
  \bibfield  {author} {\bibinfo {author} {\bibfnamefont {Alex}\ \bibnamefont
  {{Gomez-Marin}}}, \bibinfo {author} {\bibfnamefont {Tim}\ \bibnamefont
  {Schmiedl}}, \ and\ \bibinfo {author} {\bibfnamefont {Udo}\ \bibnamefont
  {Seifert}},\ }\bibfield  {title} {\enquote {\bibinfo {title} {Optimal
  protocols for minimal work processes in underdamped stochastic
  thermodynamics},}\ }\href {\doibase 10.1063/1.2948948} {\bibfield  {journal}
  {\bibinfo  {journal} {J. Chem. Phys.}\ }\textbf {\bibinfo {volume} {129}},\
  \bibinfo {pages} {024114} (\bibinfo {year} {2008})}\BibitemShut {NoStop}%
\bibitem [{\citenamefont {{Ribezzi-Crivellari}}\ and\ \citenamefont
  {Ritort}(2019)}]{ribezzi-crivellari2019}%
  \BibitemOpen
  \bibfield  {author} {\bibinfo {author} {\bibfnamefont {M.}~\bibnamefont
  {{Ribezzi-Crivellari}}}\ and\ \bibinfo {author} {\bibfnamefont
  {F.}~\bibnamefont {Ritort}},\ }\bibfield  {title} {\enquote {\bibinfo {title}
  {Large work extraction and the {{Landauer}} limit in a continuous {{Maxwell}}
  demon},}\ }\href {\doibase 10.1038/s41567-019-0481-0} {\bibfield  {journal}
  {\bibinfo  {journal} {Nat. Phys.}\ }\textbf {\bibinfo {volume} {15}},\
  \bibinfo {pages} {660--664} (\bibinfo {year} {2019})}\BibitemShut {NoStop}%
\end{thebibliography}

\end{document}